\documentclass[twocolumn,11pt]{article}
\usepackage{graphicx}

\begin{document}

\title{\bf Direct Nuclear Reactions in Lithium-Lithium Systems:\\
$^{7}$Li+$^{7}$Li at E$_{lab}$ = 2 - 16 MeV}% $^{\dagger}$}
\author{ {\bf P. Rosenthal, H. Freiesleben$^{\star}$, B. Gehrmann, I. Gotzhein,
 K. W. Potthast}\\
Institut f\"ur Experimentalphysik I, Ruhr-Universit\"at-Bochum,\\
D-44780 Bochum, Germany\\
and\\
{\bf B. Kamys, Z. Rudy}\\
Institute of Physics, Jagellonian University, PL-30059 Cracow,
Poland\\
%\vspace*{1cm}
$^{\star}$ Present address: Institut f\"ur Kern- und
Teilchenphysik,\\
 Technische Universit\"at Dresden, D-01062 Dresden,
Germany}
\maketitle

%\begin{center}
%\smallskip
%$^{\star}$ Present address: Institut f\"ur Kern- und Teilchenphysik,\\
%Technische Universit\"at Dresden, D-01062 Dresden, Germany
%\smallskip
%$^{\dagger}$ Supported by DFG under Contract No Fr 575/2-1,2,3
%\end{center}

\smallskip
\begin{tabular*}{\textwidth}{lll}
    & \parbox{0.8\textwidth}{%\begin{abstract}
%\textbf{Abstract:}
\abstract{Angular distributions of $^{7}$Li($^{7}$Li,t),
 ($^{7}$Li,$\alpha$) and ($^{7}$Li,$^{6}$He)
reactions were measured for laboratory energies from 2 - 16 MeV.
Exact finite range DWBA analyses were performed with the aim to
identify contributions of direct processes and to investigate the
applicability of DWBA to such few nucleon systems.  It turned out
that DWBA can be successfully applied to estimate differential and
total cross sections of direct transfer processes in
$^{7}$Li+$^{7}$Li interaction. The direct mechanism was found to
play a dominant role in most of these reactions but significant
contributions of other, strongly energy dependent processes were
also established. It is suggested that these processes might be due
to isolated resonances superimposed on the backround of statistical
fluctuations arising from interference of compound nucleus and
direct transfer contributions.} }& \\
  &  & \\
  & \fbox{\parbox{0.8\textwidth}{\begin{it} Nuclear Reactions: \end{it}
$^{7}$Li+$^{7}$Li at 2$\leq$ E$_{lab} \leq$ 16 MeV; measured angular
distributions and excitation functions of p, d, t, $\alpha$, and
$^{6}$He-channels; deduced direct mechanism contribution to t,
$\alpha$
and $^{6}$He-channels by DWBA analysis; reaction mechanism inferred.}} & \\
%\end{abstract} &   \\
  &                                & \\
 & {\bf PACS:} 25.70.-x; 25.70.Hi & \\
\end{tabular*}

%\smallskip
%
\clearpage  % zastêpuje \newpage w trybie "twocolumn"

\section {\label{sec:introduction} Introduction}
\hspace{1.cm}  Prominent clusterization of weakly bound lithium
nuclei may influence in various ways the mechanism of reactions in
which they take part.  A high probability of direct cluster-transfer
reactions should be a likely  consequence of clusterization,
indicated by both large spectroscopic amplitudes and small
separation energies of the clusters.  They might play a dominant
role in the system of two interacting lithium nuclei.  However,
other reaction mechanisms cannot be excluded since the small binding
energy of the entrance channel nuclei leads to a compound nucleus
with high excitation energy, i.e. with high density of states.
Hence, large compound nucleus cross sections might occur.
Furthermore, due to the clusterization, states with a specific,
simple structure may appear as isolated resonances
\cite{DOM87b,WIE93} superimposed on the statistical background of
compound nucleus decay \cite{LEI90}.
\par
The aim of the present work is twofold.  First, to study the
contribution of direct reactions in the $^{7}$Li+$^{7}$Li system
and, in turn, to estimate the magnitude of other possible
mechanisms. If direct processes were to contribute significantly we
have an opportunity to test the quality of DWBA predictions by
straightforward comparison with experimental data and, thus, to
investigate the applicability of direct reaction theory in the
extreme case of a nuclear system consisting of few nucleons.  It is
{\it a priori} not obvious whether the standard DWBA method can be
used for such a system since this approach explicitely assumes
transfer reactions to be only a perturbation to elastic scattering.
This condition may be fulfilled for transfer reaction cross sections
because they are typically smaller by an order of magnitude than
those of elastic scattering.  However, each transfer, even that of a
single nucleon, modifies the mass of target and projectile in this
light nuclear system to an extent which can hardly be considered as
perturbation.
\par
     The second intriguing aspect of the $^{7}$Li+$^{7}$Li system is the
possibility to test in DWBA calculations various optical model (OM)
potentials which give an equivalent description of elastic and
inelastic scattering data.  In a recent extensive investigation of
elastic and inelastic scattering in the $^{7}$Li+$^{7}$Li system
\cite{BAC93} no potential could be singled out on the basis of the
quality with which experimental data were described.  One may hope
that cross sections of transfer reactions will be more sensitive to
optical model potentials used to generate distorted waves for DWBA
calculations than elastic or inelastic scattering cross sections.
\par
     It seems likely that both alpha particle transfer
$^{7}$Li($^{7}$Li,t)$^{11}$B and the triton transfer
$^{7}$Li($^{7}$Li,$^{4}$He)$^{10}$Be are the best candidates for
direct - transfer reactions due to the very small separation energy
of $^{7}$Li $\rightarrow ^{4}$He + t.  These transfers, however,
significantly change the mass of target and projectile during the
collision (40 - 60$\%$). Thus, as pointed out above, the
applicability of a perturbation theory must be questioned.  In this
respect, nucleon transfer reactions, namely proton transfer
$^{7}$Li($^{7}$Li,$^{6}$He)$^{8}$Be and neutron transfer
$^{7}$Li($^{7}$Li,$^{6}$Li)$^{8}$Li, are considered most adequate.
Unfortunately, the neutron transfer reaction appears to have a
negative Q-value and therefore its cross section is very small in
the energy range studied here.  Thus only alpha particle, triton,
and proton transfers were selected for the present investigation.
They correspond to triton, alpha particle and $^{6}$He exit
channels, respectively.  The proton and the deuteron exit channels
also measured in the present experiment were used to estimate the
total fusion cross section and compound nucleus contribution to the
reactions under investigation \cite{GOT89}.
\par
     Studies of $^{7}$Li($^{7}$Li,t) and $^{7}$Li($^{7}$Li,$^{4}$He)
reactions are reported in the literature mainly for low energies
(E$_{lab}$= 2 - 6 MeV) \cite{HUB63,DZU64,CAR68} with exception of
the ($^{7}$Li,$^{4}$He) reaction for which a forward angle
excitation function was measured in the range of E$_{lab}$= 2 - 21
MeV \cite{WYB71}. The $^{7}$Li($^{7}$Li,$^{4}$He)$^{10}$Be reaction
was also studied at 26 and 30 MeV with the aim to look for resonant
states of $^{10}$Be \cite{GLU71}, while the
$^{7}$Li($^{7}$Li,$^{11}$B)t reaction was investigated at 79.6 MeV
as part of a study of reactions leading to multi - neutron final
states \cite{CER74}. All these investigations were not concerned
with the questions asked here.
\par
     The $^{7}$Li($^{7}$Li,$^{6}$He) reaction was reported in the literature
at very low energies (E$_{lab}$= 3 MeV \cite{CAR64}, E$_{lab}$=
3-3.8 MeV \cite{STR68})  where only spectra at small reaction angles
were measured, and at energies above our energy range (Bochkarev et
al. \cite{BOC88}, E$_{lab}$=22 MeV). Angular distributions of
transitions to the ground and the first excited state of $^{8}$Be
were found in the latter experiment to show pronounced oscillations
and to be rather steep.  The authors suggested a direct reaction
mechanism since DWBA calculations agreed reasonably well with the
experimental data.  Moreover, estimation of the compound nucleus
contribution made by means of the Hauser-Feshbach model indicated
\cite{BOC88} that the compound nucleus mechanism is responsible for
only a small part (approximately 10$\%$) of the experimental cross
section.  Thus the $^{7}$Li($^{7}$Li,$^{6}$He)$^{8}$Be reaction may
be used to study the applicability of the DWBA formalism for this
few - nucleon system and for testing various OM potentials in the
$^{7}$Li+$^{7}$Li system.
\par
In the present work angular distributions of light ejectiles
(tritons and $^{4}$He) are measured in the energy range from
E$_{lab}$= 2 - 16 MeV while those for $^{6}$He are restricted to 8 -
16 MeV. The experimental procedure and results are given in the next
chapter. The analysis of the data in the frame of the DWBA formalism
is presented in the third chapter while a summary with conclusions
is provided in the last one.

\section{\label{sec:experiment} Experimental procedure}
\hspace{1.cm}

The experiments were carried out at the 4 MV Dynamitron Tandem
accelerator at the Ruhr-Universit\"at Bochum. $^7$Li-beams were
produced with a deflection sputter source; for these experiments the
beam currents (max. 2.5 $\mu$A Li$^-$ ions) were limited to 30 nA up
to 100 nA (electric) in order not to destroy the targets and
backings. The beam was focused onto the targets via two collimators
($\oslash$ =1.5 mm) 40 and 60 cm upstream of the target.  Targets of
metallic lithium in natural isotopic abundance (92.5\% $^7$Li, 7.5\%
$^6$Li) were prepared by evaporation onto different backing
materials adapted to each experimental setup.

\par
\begin{figure}[ht!]
  \begin{center}
  \includegraphics[width=0.5\textwidth]{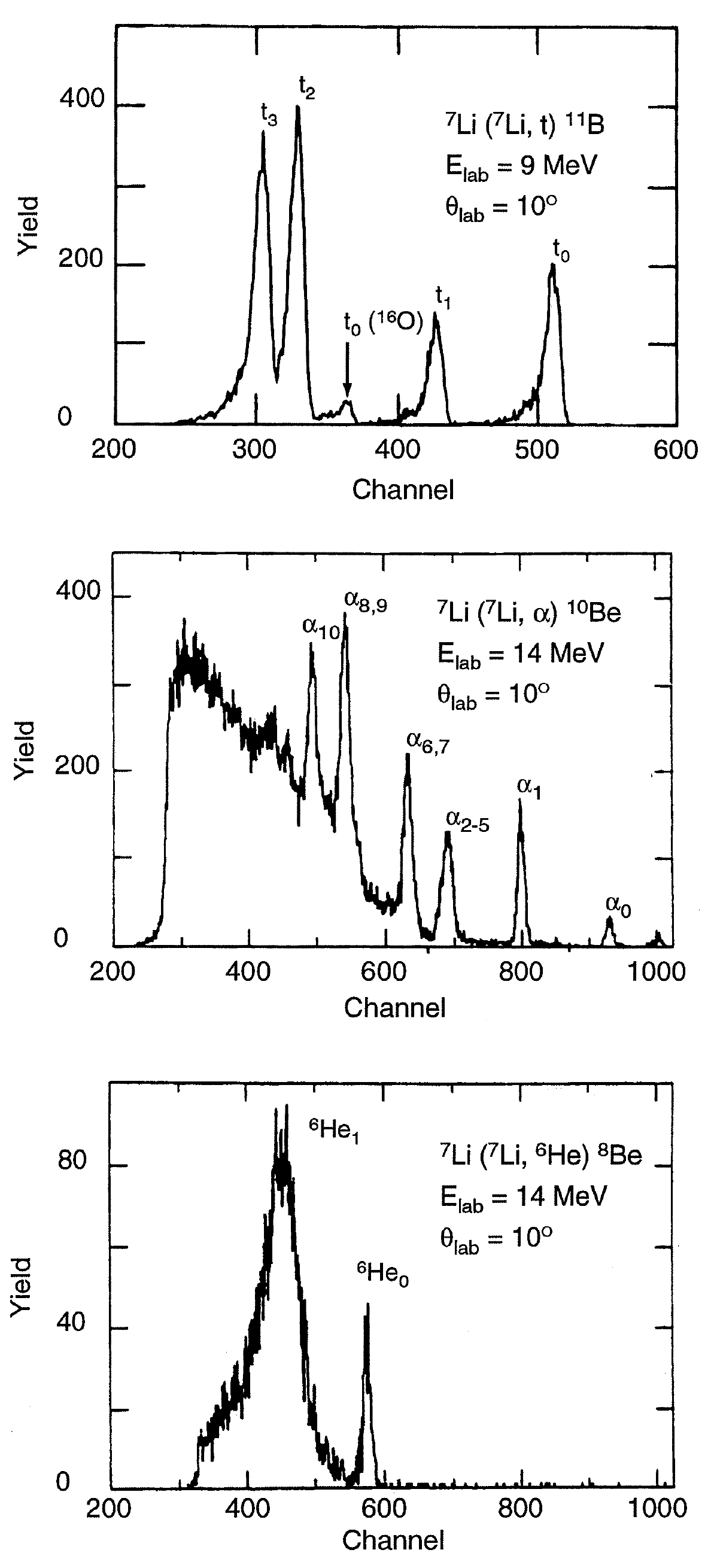}\\
  \caption{Experimental spectra of tritons ($\theta_{lab}=30^{\circ}$,
         E$_{lab}$=9 MeV), $\alpha$-particles ($\theta_{lab}=10^{\circ}$,
         E$_{lab}$=8 MeV) and $^{6}$He ejectiles ($\theta_{lab}=10^{\circ}$,
         E$_{lab}$=16 MeV)}\label{fig:F01}
  \end{center}
\end{figure}

     The standard technique to identify charged particles in low energy nuclear
reactions is the $\Delta$E-E-discrimination. This technique was
employed with triple telescopes of surface barrier detectors in
order to cover the broad dynamical range of the p, d, t, $\alpha$
and $^6$He exit channels in one measurement. These exit channels
were investigated in an energy range from E$_{lab}$ = 2 - 16 MeV in
steps of 0.5 MeV by measuring the differential cross sections from
$\theta_{lab}$ = 0$^\circ$ - 80$^\circ$.  Some typical spectra are
depicted in Fig. \ref{fig:F01}.

%----------------------------------------------------------------------------------

%---------------------------------------------------------------------------------------

Due to the identity of
projectile and target angular distributions are symmetrical with
respect to
90$^{\circ}$ in the center - of - mass system. Hence, it is sufficient
to measure up to 80$^{\circ}$ in the laboratory system in order to obtain
the whole angular distribution.

\par
     To determine absolute cross sections for these exit channels three
separate experimental setups were utilized:
\par
     1. With the first differential cross sections were measured from
$\theta_{lab}$ = 10$^\circ$ - 80$^\circ$ in steps of 10$^\circ$.
These measurements were carried out with two triple telescopes.
The first telescope with an aperture of 0.311 msr and total
energy resolution of 280 keV covered the forward angles from
$\theta_{lab}$ = 10$^\circ$ up to 40$^\circ$. The second telescope
(aperture: 0.179 msr; total energy resolution: 240 keV) was
used to cover the more backward angles from $\theta_{lab}$ = 50$^\circ$
up to 80$^\circ$.
In front of each telescope aluminium foils were positioned to
absorb $^7$Li ions elastically scattered from the Ni-backing.
A thickness from 30 up to 55 $\mu$m was sufficient depending on
beam energy and laboratory angle of the telescopes was sufficient.
For this setup transmission targets were used consisting of metallic
$^7$Li with an area density of 55 $\mu$g / cm$^2$ evaporated onto a
Ni-backing with an area density of 90 $\mu$g / cm$^2$.
\par
     2. The differential cross section of the $^{7}$Li($^{7}$Li,
$\alpha_{0,1}$)$^{10}$Be reaction at 0$^\circ$ and 5$^\circ$
was measured separately with a special target.  It consisted of 55
$\mu$g / cm$^2$ metallic $^{7}$Li evaporated onto Ni-foils; their
thickness of 5.0 $\mu$m up to 35.4 $\mu$m, depending on beam energy,
sufficed to fully stop the $^7$Li-beam, rendering possible
measurements at 0$^\circ$.  One triple telescope with an aperture of
0.314 msr was utilized.  Both measurements were carried out
independently on absolute scale.  The purpose of this second
experiment was twofold.  First, to extend the angular distributions
to 0$^\circ$, a region which is very sensitive to transfers with
orbital angular momentum of {\it l}=0.  Second, to verify the
experimental data of Wyborny and Carlson \cite{WYB71} who found very
prominent structures in the 0$^{\circ}$ - excitation function (cf.
Fig. \ref{fig:F04}).
\par
     3. For either setup the target thickness was determined via
Rutherford scattering of $^{58}$Ni$^{4+}$ off $^{7}$Li in a
well defined geometry.  Experimental details of target preparation
and thickness determination may be found elsewhere \cite{DOM87a}.
\par
\begin{figure*}[ht!]
  \begin{center}
  \includegraphics[width=0.9\textwidth]{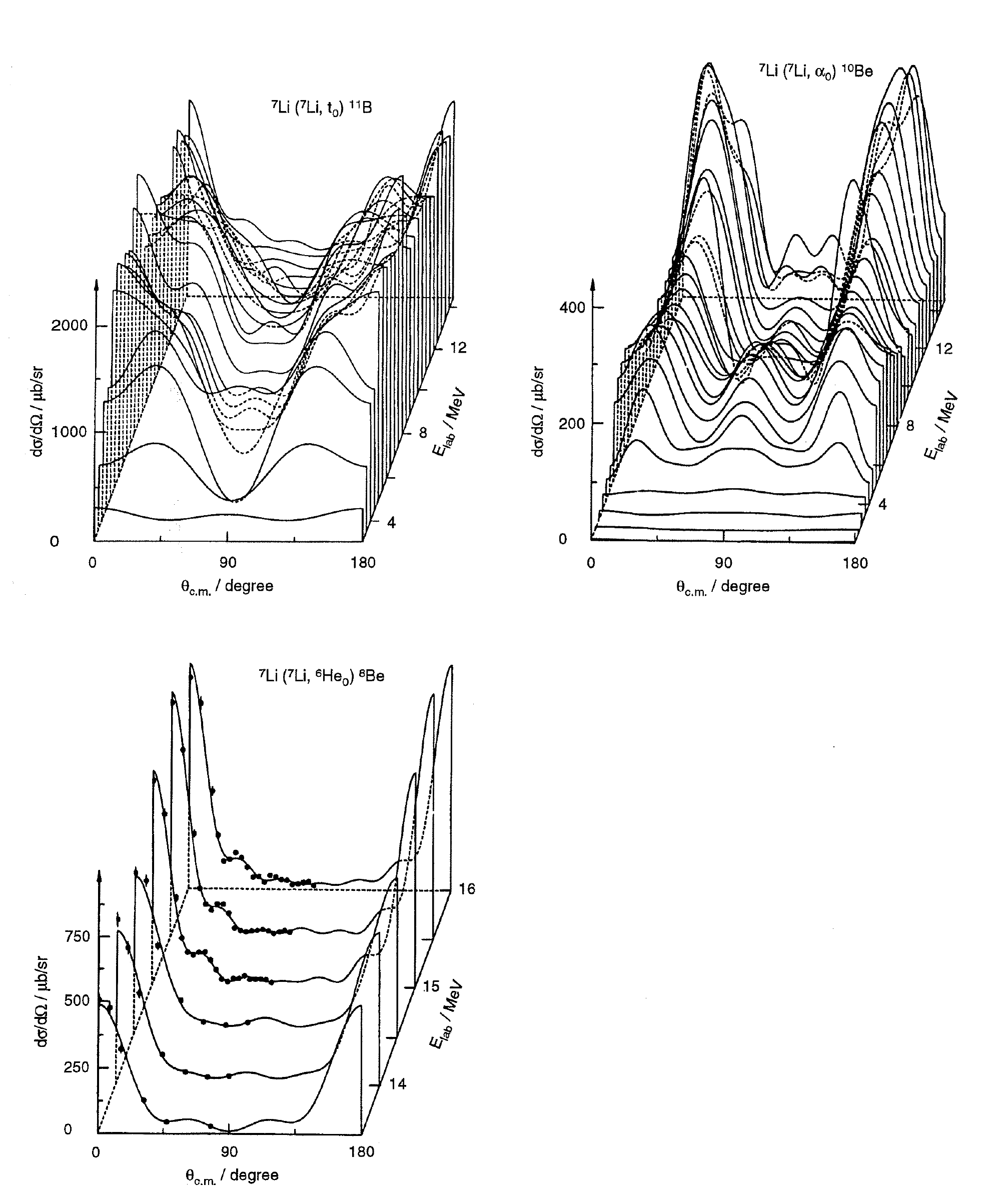}\\
  \caption{Overview of angular distributions of ground state
         transitions $^{7}$Li($^{7}$Li,t$_{0}$),
         $^{7}$Li($^{7}$Li,$\alpha_{0}$) and
         $^{7}$Li($^{7}$Li,$^{6}$He$_{0}$) represented by result of a least
         square fit of series of Legendre polynomials to the data.
         For reason of legibility experimental data are only included for
         the $^{6}$He channel.}\label{fig:F02}
  \end{center}
\end{figure*}

     It can be inferred from Fig. \ref{fig:F01} that cross sections for triton exit
channels are readily determinable for transitions to the ground
state and the first three excited states in $^{11}$B.  In case of
the $\alpha$-particle exit channel only transitions to the ground
and first excited state were evaluated, and for the $^{6}$He-exit
channel only the ground state transition was selected for the
analysis beause the higher lying excited states in either case are
residing on a continuous background of three particle decays, namely
$^{14}$C$^{\star} \rightarrow ^{9}$Be + n + $\alpha$ and
$^{14}$C$^{\star} \rightarrow \alpha + ^{10}$Be$ \rightarrow
^{6}$He$ + \alpha$, respectively, the shape of which is unknown,
rendering rather difficult a reliable evaluation.  In Fig.
\ref{fig:F02} we present an overview of angular distributions for
the three ground state transitions investigated.  They are
represented by a least square fit of a Legendre polynomial expansion
to the data.  For reason of legibility experimental data are only
included for the $^{6}$He-channel.
Experimental uncertainties are of symbol size, if not shown
explicitely.

%-----------------------------------------------------------------
%--------------------------------------------------------------------------

\begin{figure}[t!]
  \begin{center}
  \includegraphics[width=0.5\textwidth]{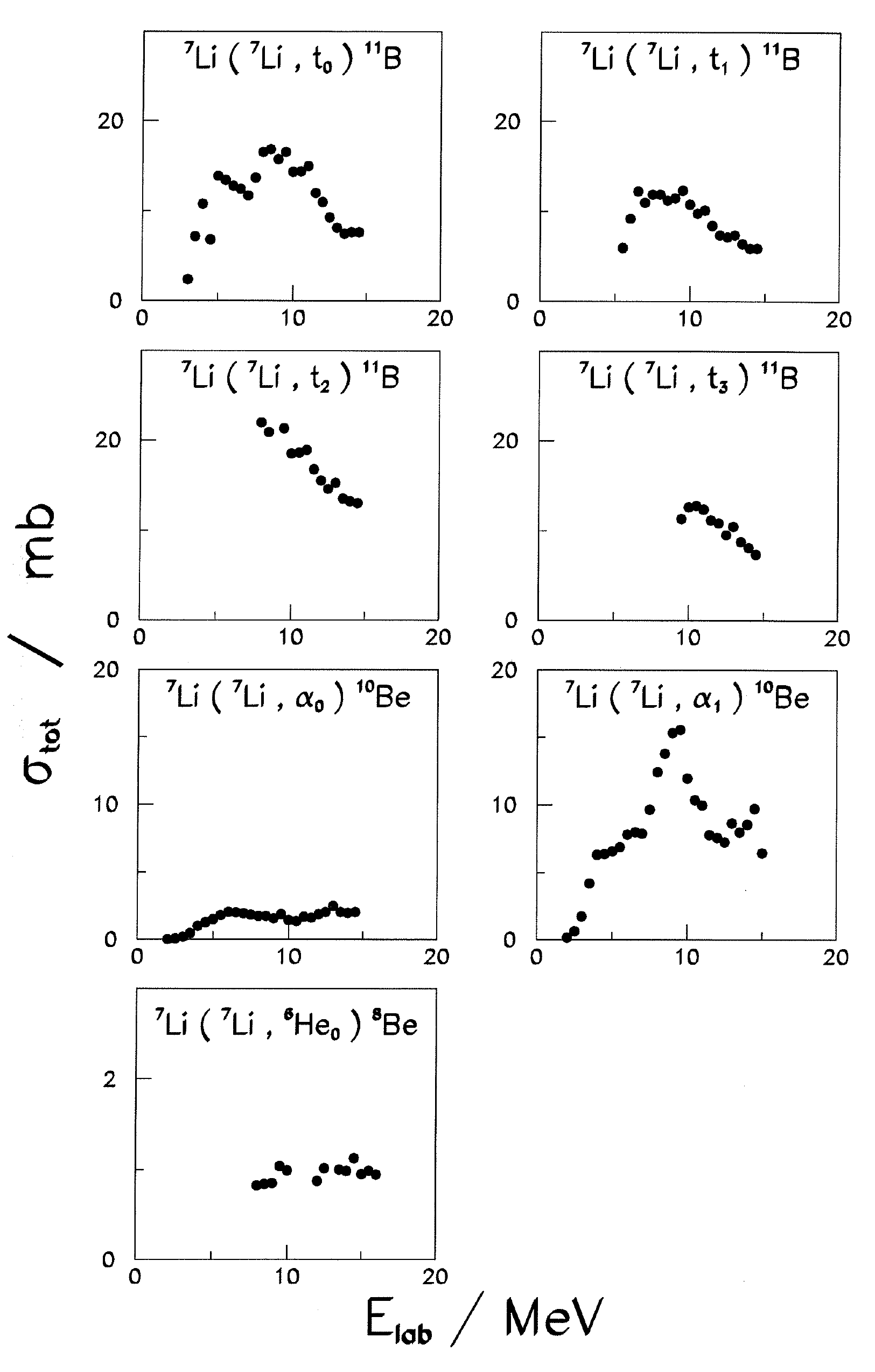}\\
  \caption{Excitation functions of angle integrated cross sections
         for $^{7}$Li($^{7}$Li,t)$^{11}$B to the four lowest states of
         $^{11}$B; (g.s.;3/2$^{-}$), (2.125 MeV;1/2$^{-}$),
         (4.445 MeV;5/2$^{-}$) and
         (5.020 MeV;3/2$^{-}$); excitation functions of angle integrated
         cross sections for $^{7}$Li($^{7}$Li,$^{4}$He)$^{10}$Be
         to the two lowest
         states of $^{10}$Be; (g.s.;0$^{+}$) and (3.368 MeV;2$^{+}$)
         and excitation function of angle integrated cross section for
         $^{7}$Li($^{7}$Li,$^{6}$He)$^{8}$Be$_{g.s.}$.}\label{fig:F03}
  \end{center}
\end{figure}

\begin{figure}[ht!]
  \begin{center}
  \includegraphics[width=0.5\textwidth]{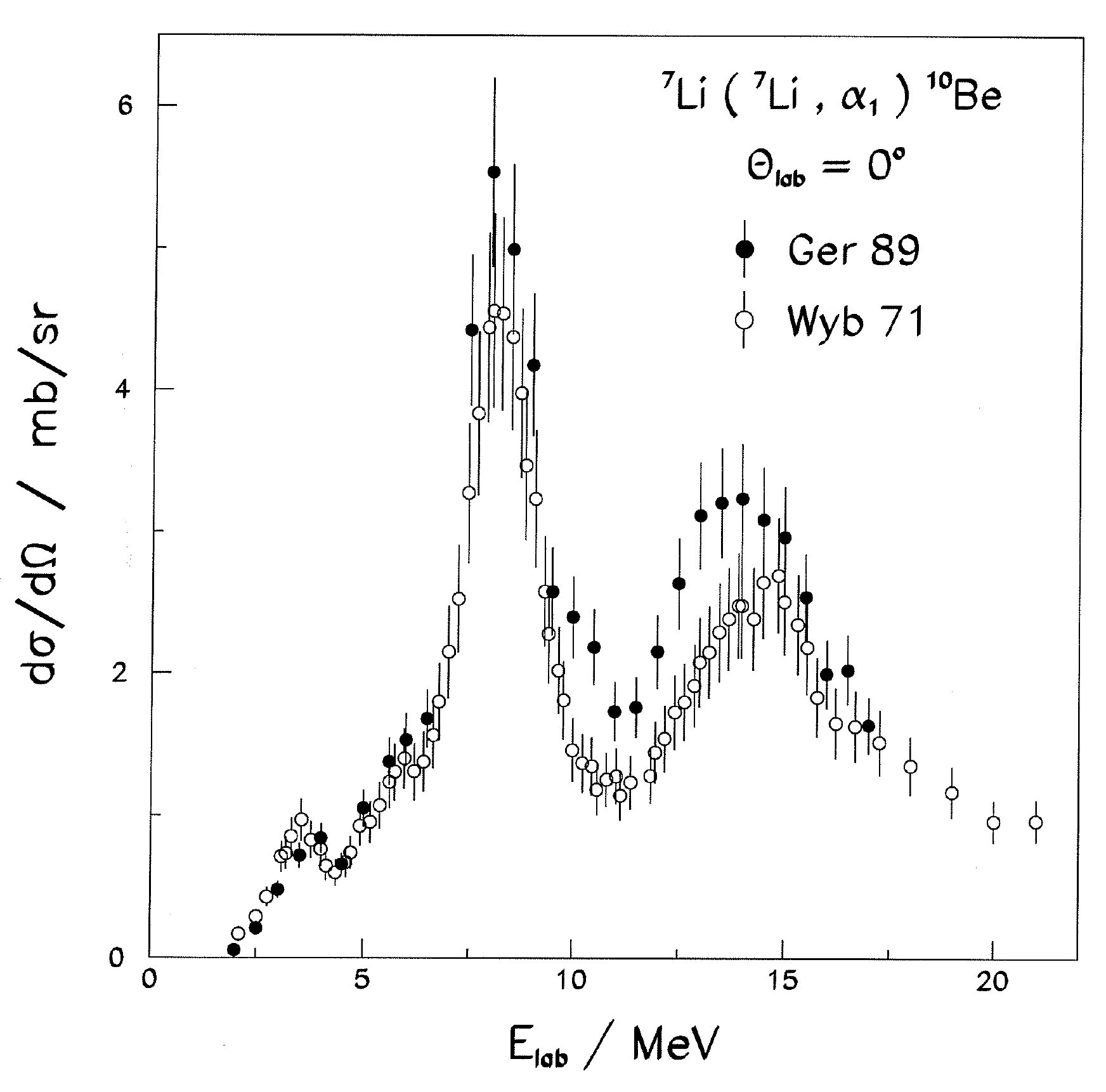}\\
  \caption{Excitation function for $^{7}$Li($^{7}$Li,$\alpha_{1}$) measured at
         $\theta_{lab}=0^{\circ}$ (full dots) together with data of Wyborny and
         Carlson (open dots) \cite{WYB71}}\label{fig:F04}
  \end{center}
\end{figure}

 Fig. \ref{fig:F03} contains the excitation functions of angle
integrated cross sections for the triton channel (four lowest states
of $^{11}$B), the $\alpha$-channel (two lowest states of $^{10}$Be)
and the $^{6}$He-channel (ground state of $^{8}$Be).

%----------------------------------------------------------------------
%----------------------------------------------------------------------------------

Fig. \ref{fig:F04} shows the excitation function for
$^{7}$Li($^{7}$Li,$\alpha_{1}$) measured at 0$^\circ$ together with
the data of Wyborny and Carlson \cite{WYB71}.  A very good agreement
can be stated.

\section {\label{sec:DWBA} DWBA analysis}

The calculations were performed using the exact finite
range DWBA computer code Jupiter-5 \cite{TAM80} in both
representations ({\it prior} and {\it post}) of the transition
potentials.  In the ideal case of the exact knowledge of
transition potentials e.g. from some microscopic model,
DWBA calculations should lead to the same result in both
representations \cite{SAT83}.  Therefore the quality of agreement
between results of calculations made within either
representations may be taken as indication of both the
applicability of the DWBA formalism in its standard form and
the proper choice of potentials.

%------------------------------------------------------------------------------
%-------------------------------------------------------------------------------

\par
     We followed the standard prescription of DWBA for choosing
the strong interaction potential responsible for the transfer. This
is discussed in the following using the proton transfer reaction
$^{7}$Li($^{7}$Li,$^{6}$He)$^{8}$Be as an example. Either the
potential which binds the proton to $^{7}$Li ({\it post}) or that
which binds the proton to $^{6}$He ({\it prior} representation) was
taken as nuclear transition potential.

%-------------------------------------------------------------------
%-------------------------------------------------------------------
\begin{table*}[ht!]
\begin{center}
\caption{\label{tab:1} Optical model potentials used in the DWBA
calculations.}
\begin{tabular}{|cc|r|c|c|r|c|c|c|c|}
\hline System & Family& U & r$_{U}$ & a$_{U}$ & W & r$_{W}$ &
a$_{W}$ & r$_{C}$ &
 Ref. \\
{\mbox{} } &{\mbox{} } & MeV &  fm  &  fm &  MeV  &  fm  &  fm  & fm
&
{\mbox{} }  \\
\hline \hline $^{7}$Li+$^{7}$Li & S-1 &   3 & 1.456 & 1.416 & 9.8 &
1.369 & 0.441 & 1.25 &
\cite{BAC93} \\
{\mbox{} }      & S-2 &  21 & 0.802 & 1.279 & 13.8 & 1.354 & 0.409 &
1.25 &
\cite{BAC93} \\
{\mbox{} }      & S-3 &  49 & 0.800 & 1.036 & 18.2 & 1.354 & 0.387 &
1.25 &
\cite{BAC93} \\
{\mbox{} }      & S-4 &  69 & 0.940 & 0.860 & 21.2 & 1.381 & 0.345 &
1.25 &
\cite{BAC93} \\
{\mbox{} }      & S-5 & 101 & 0.965 & 0.781 & 23.3 & 1.397 & 0.320 &
1.25 &
\cite{BAC93} \\
{\mbox{} }      & S-6 & 136 & 0.994 & 0.721 & 26.2 & 1.411 & 0.296 &
1.25 &
\cite{BAC93} \\
\hline $^{10}$Be+$^{4}$He & S  & 136 & 0.673 & 0.792 & 35.6 & 0.924
& 0.137 & 0.795&
\cite{ENG77} \\
\hline $^{11}$B + t     & V  & 133 & 0.923 & 0.570 & 19.5 & 1.090 &
0.220 & 0.923&
\cite{HAR80} \\
\hline
\end{tabular}
\end{center}
\begin{footnotesize}
Real parts of all potentials have the Woods-Saxon form with the
following  parametrization  of radii : R=r$_0$*(A$_{1}^{1/3} +
A_{2}^{1/3}$).  The imaginary potentials of the "V" families have
the volume shape of Woods-Saxon form while those of the "S" families
use the surface shape
of derivative of Woods-Saxon form. \\
\end{footnotesize}
\end{table*}

Thus it was assumed that perfect cancellation occurs of the so
called "indirect transition potentials" i.e. the core-core
($^{7}$Li-$^{6}$He) interaction potential is equivalent to either
the optical model (OM) potential for the $^{7}$Li-$^{7}$Li channel
({\it prior} representation) or the optical model potential for the
$^{6}$He-$^{8}$Be channel ({\it post} representation).  Such an
assumption seems to be justified because, in contradistinction to a
rather good knowledge of entrance/exit channel optical model
potentials, the core-core potential is not known and it is necessary
to make assumptions concerning this potential.  The best
approximation should be an optical model potential for scattering of
the core - core nuclear system.  However, this potential is usually
not known, and moreover, it is not obvious at which energy of the
relative motion of the core - core system such potential should be
taken.  Thus the standard prescription for choosing the nuclear
transition potential is to approximate the core - core potential by
the optical model potential of either the entrance or the exit
channel.

\par
     The binding potentials were taken in Woods-Saxon form.
Their geometrical parameters were arbitrarily fixed at the following
values: the reduced radius r$_{0}$ = 0.97 fm (for symmetrical
parametrization i.e. R=r$_{0}*$(A$_{core}^{1/3}$ +
A$_{cluster}^{1/3}$) and the diffuseness parameter a= 0.65 fm.  Such
values were successfully used when describing the alpha particle
transfer in $^{14}$N(d,$^{6}$Li)$^{10}$B reactions \cite{OEL79}. The
depth of the potentials was fitted to reproduce the appropriate
binding energy.
\par
     The Coulomb "indirect transition potentials" were all
taken into account in the present analysis and they were used in the
standard form of uniformly charged spheres.
\par
     The Jupiter-5 computer code evaluates the reaction
amplitudes for one orbital of the cluster in the donor and one
orbital in the acceptor nucleus.  For some reactions e.g. alpha
particle transfers to $^{11}$B$_{g.s.}$, $^{11}$B$_{4.45}$ and
$^{11}$B$_{5.02}$ two reaction amplitudes must be calculated because
of the two different orbitals of an alpha particle cluster in the
boron nucleus.  The coherent superposition was performed by means of
a separate computer code SQSYM \cite{KAM88} which also took care of
the antisymmetrization of amplitudes which is required by the
identity of projectile and target.
\par
     The introduction of free parameters was avoided by the
following procedure:  The transition potentials were fixed according
to the standard prescription described above.

%------------------------------------------------------------------
%------------------------------------------------------------------
\begin{table*}[ht!]
\begin{center}
\caption{\label{tab:2} Cluster spectroscopic amplitudes used in the
DWBA calculations.}
\begin{tabular}{|r|c|c|c|c|c|c|c|}
\hline
 Nucleus & Core & Cluster &  n  &  l  &  j  &  C*A    &   Ref.  \\
\hline \hline $^{7}Li_{g.s.} (3/2^{-})$&  t      & $^{4}He$  &  1  &
1  &  1   &  1.084  &
\cite{KWA89,KUR73}  \\
{\mbox{} }               &$^{4}He$ &     t     &  1  &  1  &  3/2 &
1.084  &
\cite{KWA89,KUR73}  \\
{\mbox{} }               &$^{6}He$ &     p     &  0  &  1  &  3/2 &
-0.831  &
\cite{KWA87}  \\
\hline $^{8}Be_{g.s.} (0^{+})$  &$^{7}Li$ &     p     &  0  &  1  &
3/2 &  1.287  &
\cite{KWA87}  \\
\hline $^{10}Be_{g.s.}(0^{+})$  &$^{7}Li$ &     t     &  1  &  1  &
3/2 &  0.556  &
\cite{KUR73}  \\
\hline $^{10}Be_{3.37}(2^{+})$  &$^{7}Li$ &     t     &  1  &  1  &
1/2 &  0.568  &
\cite{KUR73}  \\
{\mbox{} }              &{\mbox{} }&{\mbox{} } &  1  &  1  &  3/2 &
0.040  &
\cite{KUR73}  \\
{\mbox{} }              &{\mbox{} }&{\mbox{} } &  0  &  3  &  5/2 &
0.604  &
\cite{KUR73}  \\
{\mbox{} }              &{\mbox{} }&{\mbox{} } &  0  &  3  &  7/2 &
-0.299  &
\cite{KUR73}  \\
\hline $^{11}B_{g.s.}(3/2^{-})$& $^{7}Li$ & $^{4}He$  &  2  &  0  &
0  & -0.509  &
\cite{KWA86a}  \\
{\mbox{} }              &{\mbox{} }&{\mbox{} } &  1  &  2  &   2  &
0.629  &
\cite{KWA86a}  \\
\hline $^{11}B_{2.13}(1/2^{-})$& $^{7}Li$ & $^{4}He$  &  1  &  2  &
2  & -0.585  &
\cite{KWA86a}  \\
\hline $^{11}B_{4.45}(5/2^{-})$& $^{7}Li$ & $^{4}He$  &  1  &  2  &
2  &  0.725  &
\cite{KWA86}  \\
{\mbox{} }              &{\mbox{} }&{\mbox{} } &  0  &  4  &   4  &
0.018  &
\cite{KWA86}  \\
\hline $^{11}B_{5.02}(3/2^{-})$& $^{7}Li$ & $^{4}He$  &  2  &  0  &
0  & -0.292  &
\cite{KWA86}  \\
{\mbox{} }              &{\mbox{} }&{\mbox{} } &  1  &  2  &   2  &
-0.322  &
\cite{KWA86}  \\
\hline
\end{tabular}
\end{center}
\begin{footnotesize}
"n" corresponds to the number of nodes of the bound
state radial wave function (exluding r=0 and infinity), \\
"l" is the orbital angular momentum
of relative motion of the cluster and the core, \\
"j" is the total angular momentum of the orbital, \\
"C*A" denotes the product of the isospin
Clebsch-Gordan coefficient and the spectroscopic amplitude. \\
\end{footnotesize}
\end{table*}
%---------------------------------------------------------------------
%---------------------------------------------------------------------

%
Parameters of optical model potentials were taken from the
literature, whenever it was possible i.e. for $^{7}$Li+$^{7}$Li
\cite{BAC93}, $^{10}$Be+$\alpha$ \cite{ENG77} and $^{11}$B+t
\cite{HAR80} systems, or they were approximated by $^{7}$Li+$^{7}$Li
potentials for the unaccessible elastic $^{8}$Be+$^{6}$He channel.
All potential parameters are listed in Table \ref{tab:1}.
The spectroscopic amplitudes were taken from shell model
calculations published in the literature for protons \cite{KWA87},
tritons \cite{KUR75,KWA89}, and for $\alpha$ - particles
\cite{KUR73,KWA86,KWA86a}.
Their values are given in Table \ref{tab:2}.
In that sense all calculations of cross sections were
performed without any free parameter.

\smallskip

\subsection {\label{sec:he6} The proton transfer reaction -
$^{7}$Li($^{7}$Li,$^{6}$He)$^{8}$Be }

\smallskip
\par
     A qualitative inspection of the experimental data suggests
the ($^{7}$Li,$^{6}$He) reaction to be dominated by a direct reaction
mechanism.  The differential cross section of this reaction is
significantly smaller (at least two orders of magnitude) than
the elastic scattering cross section in the $^{7}$Li+$^{7}$Li system
\cite{BAC93} compared at corresponding scattering angles.
Therefore one is allowed to treat the proton transfer
reaction ($^{7}$Li,$^{6}$He) as a perturbation to the elastic scattering
and hence to apply the distorted wave Born approximation.
Furthermore, the transfer of one nucleon results in a
relatively small rearrangement between the interacting
lithium nuclei and may thus, also from this point of view, be
considered as a perturbation.  These arguments led us to
consider the $^{7}$Li($^{7}$Li,$^{6}$He)$^{8}$Be reaction as the best candidate
among the reactions under investigation for testing the
applicability of DWBA to such light nuclear system.
\par
    Calculations of angular distributions for proton transfer
were performed for 8, 10, 12, 14 and 16 MeV laboratory energy for
which experimental data were measured in the present work.
Additional calculations were performed at an energy of 22 MeV for
the reaction leading to both the ground and the first excited state
of $^{8}$Be rendering possible a comparison with the experimental
data of Bochkarev et al. \cite{BOC88}.

%----------------------------------------------------------------------
\begin{figure}[ht!]
  \begin{center}
  \includegraphics[width=0.49\textwidth]{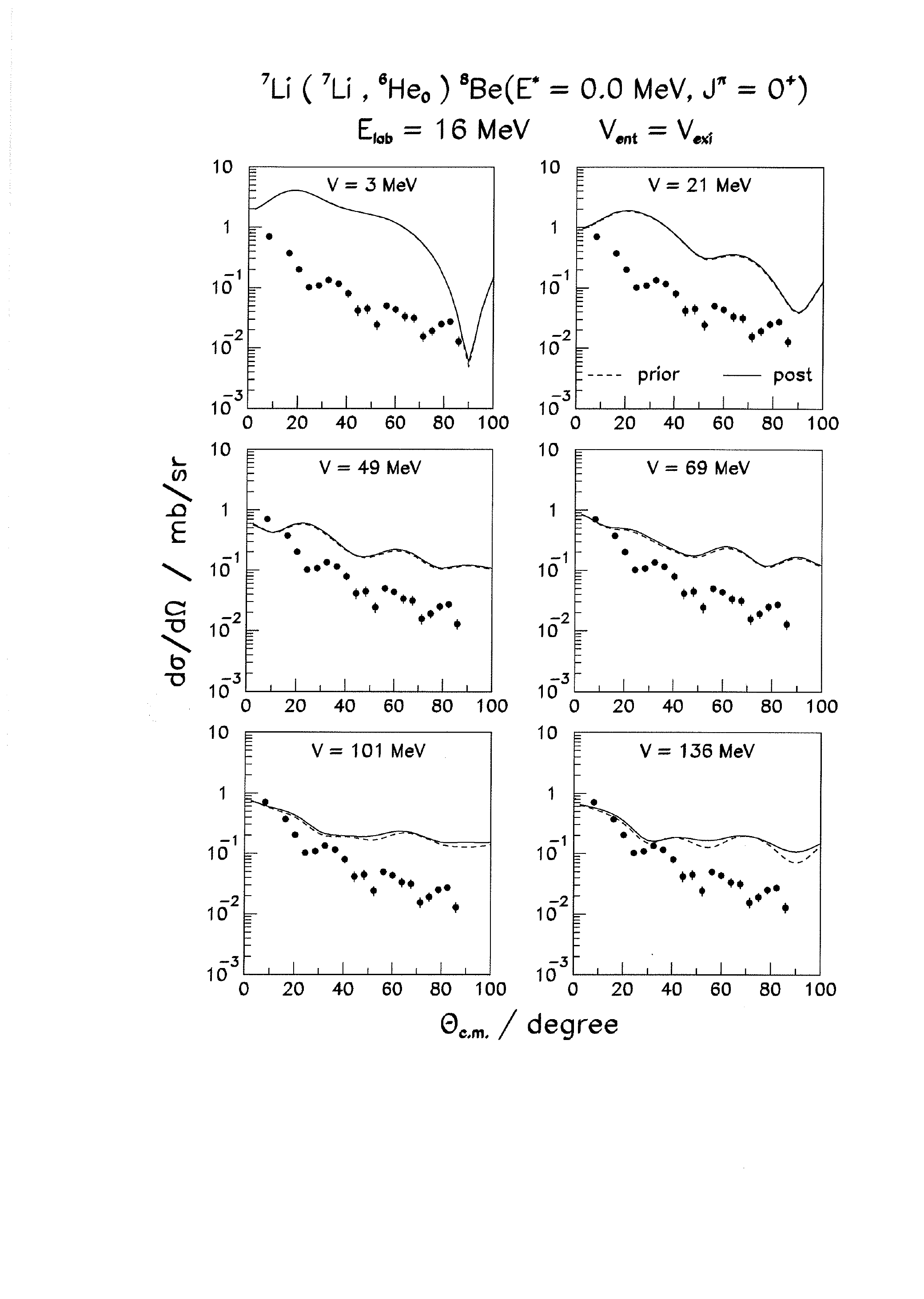}\\
  \caption{Experimental angular distribution of
          $^{7}$Li($^{7}$Li,$^{6}$He)$^{8}$Be$_{g.s.}$ reaction at
          E$_{lab}$=16 MeV (full
          dots).  Solid and dashed curves represent results of
          DWBA calculations in {\it post} and {\it prior}
          representation, respectively.  Six OM potentials were
          applied which describe elastic scattering in the $^{7}$Li+$^{7}$Li
          system  equally well \cite{BAC93}.  The "family" of the OM
          potential parameters is specified by quoting the
          depth of its real part.  The same OM potentials were
          used in entrance and exit channel.}\label{fig:F05}
  \end{center}
\end{figure}
%--------------------------------------------------------------------------

%
\par
\begin{figure}[ht!]
  \begin{center}
  \includegraphics[width=0.5\textwidth]{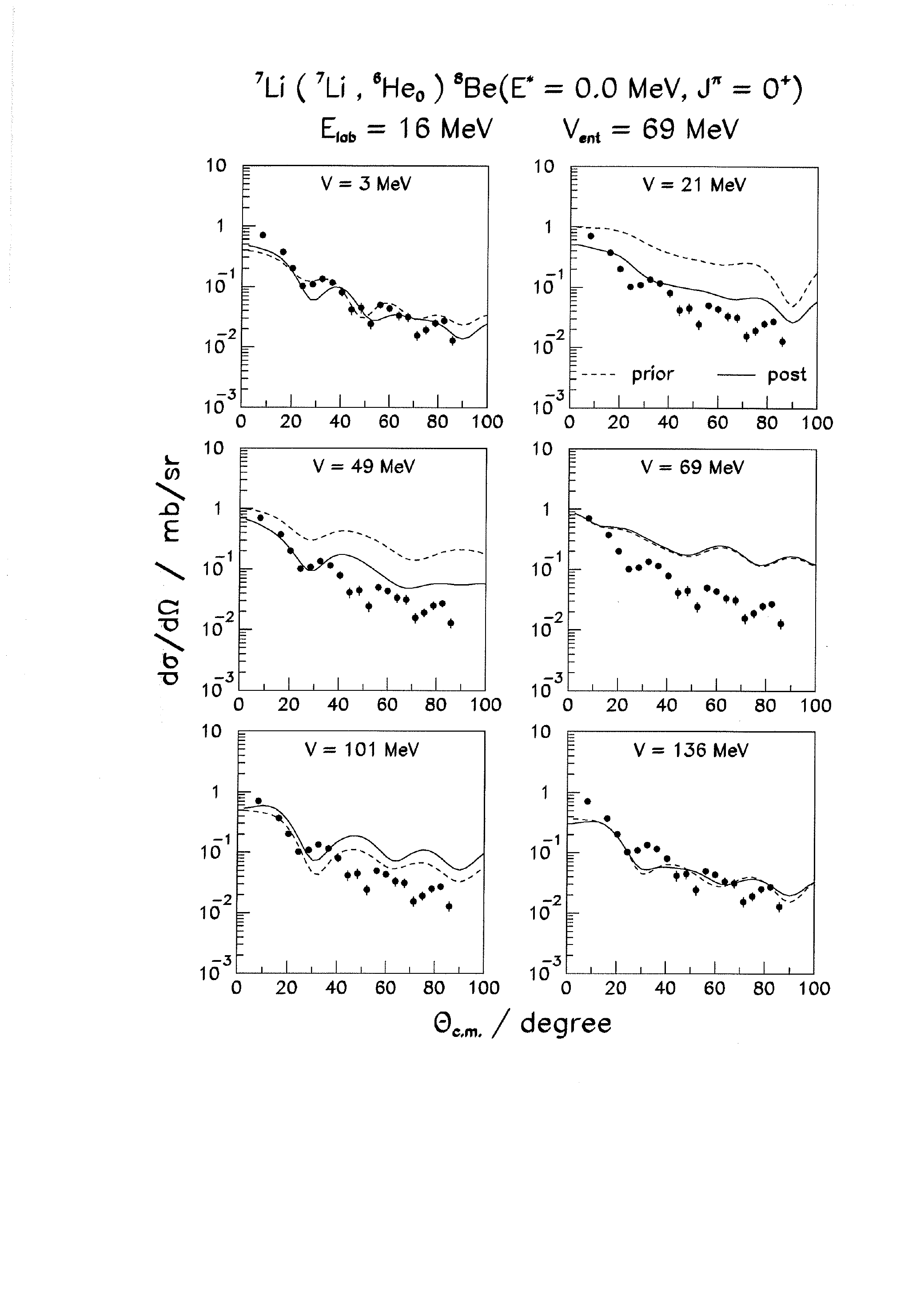}\\
  \caption{Same as Fig. \ref{fig:F05}, but the entrance channel OM potential
          is fixed (family \#4, i.e. "69 MeV" of \cite{BAC93}) while
          all six equivalent potentials of ref. \cite{BAC93} are
          used for the exit channel.}\label{fig:F06}
  \end{center}
\end{figure}
\begin{figure}[ht!]
  \begin{center}
  \includegraphics[width=0.5\textwidth]{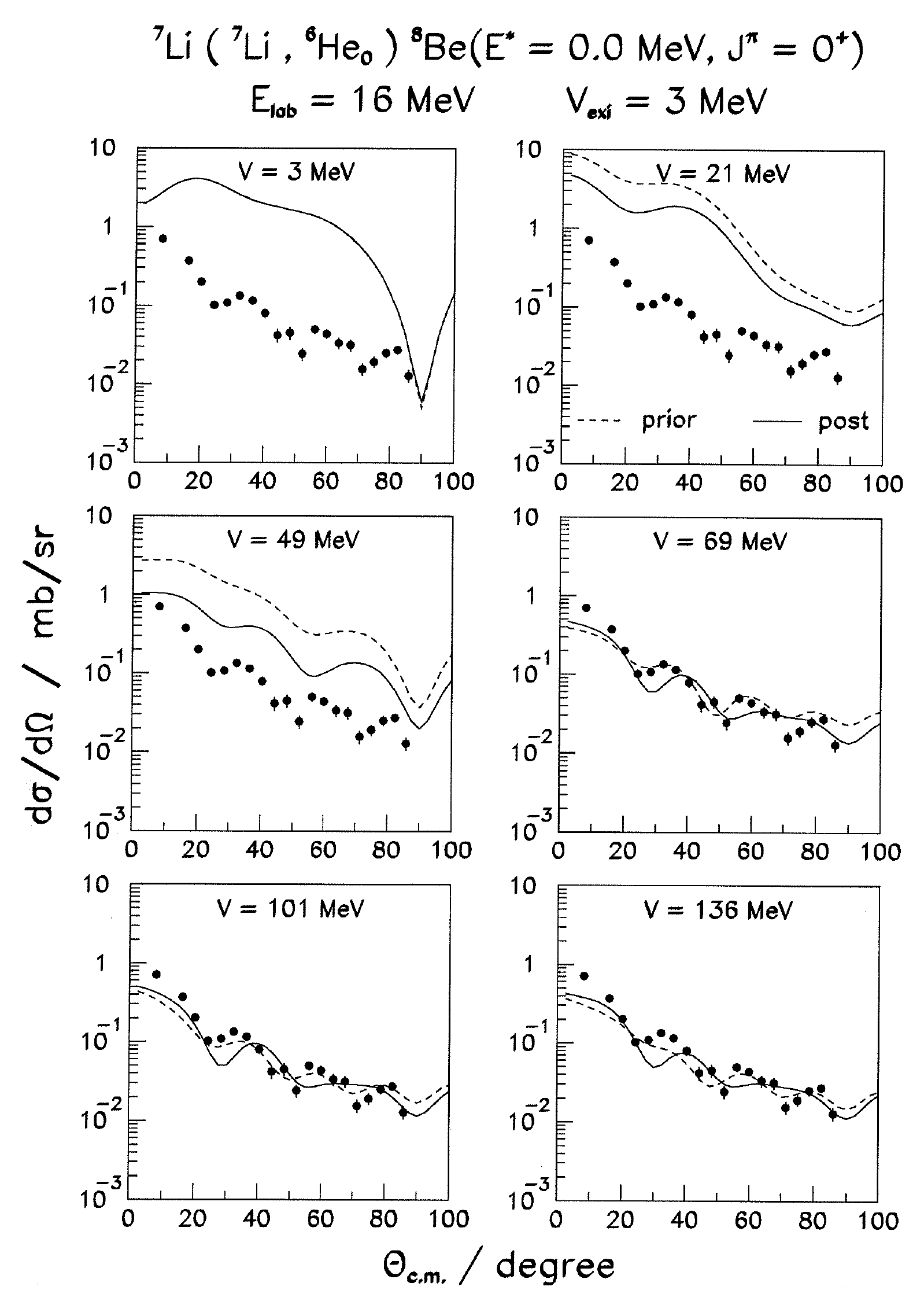}\\
  \caption{Same as Fig. \ref{fig:F05}, but the exit channel OM potential is
          fixed (family \#1, i.e. "3 MeV" of \cite{BAC93} while all
          six equivalent potentials of ref. \cite{BAC93} are used
          for the exit channel.  Note, that for the most
          shallow potential in the entrance channel results of
          calculations in {\it prior} and {\it post} representations
          are almost indistinguishable.}\label{fig:F07}
  \end{center}
\end{figure}

     The quality of reproduction of the experimental data, the
similarity of results in {\it prior} and {\it post} representations,
and the dependence of the results on the optical model potentials
was found to be almost the same for the energies under
investigation.  Therefore, we present in Figs. \ref{fig:F05},
\ref{fig:F06} and \ref{fig:F07} only results for one bombarding
energy,i.e., 16 MeV.

%-----------------------------------------------------------------
%--------------------------------------------------------------------

%
\par
     Identical optical model potentials used for both entrance
and exit channel lead to a perfect equivalence of the {\it prior}
and {\it post} representations.  This can be inferred from Fig.
\ref{fig:F05} where angular distributions evaluated in {\it prior}
(dashed lines) and in {\it post} representation (full lines) almost
coincide for six different "families" of parameters of OM potentials
listed in Table \ref{tab:1}.  Thus we conclude, that DWBA is
applicable for this reaction under the stated conditions.
%

%----------------------------------------------------------------------------
%----------------------------------------------------------------------

%
\par
It remains an open question whether the
$^{7}$Li($^{7}$Li,$^{6}$He)$^{8}$Be reaction is realized in nature
only via a direct mechanism. It may be answered by comparing the
experimental angular distributions with the theoretical ones.  It is
visible in Fig. \ref{fig:F05} that theoretical cross sections
surpass the experimental data if the calculations are performed with
the shallow optical model potentials.  This may hint at the
possibility to select some potentials among those which reproduce
equally well the experimental elastic scattering data.  However, in
order to do so one has to vary the entrance and exit channel OM
potentials independently since there is no obvious reason to assume
them to be equal.  The calculations were thus repeated for all
possible pairs of entrance/exit channel potentials listed in Table
\ref{tab:1}.  Figs. \ref{fig:F06} and \ref{fig:F07} illustrate part
of the results obtained.

\par
     In Fig. \ref{fig:F06} results of calculations are shown obtained with
a rather deep potential in the entrance channel ("69 MeV" family)
and with various potentials for the exit channel. A reasonable
reproduction of the data is possible only for a very shallow
potential ("3 MeV" family) in the exit channel. It indicates that in
spite of the small difference in mass partition ($^{7}$Li+$^{7}$Li
vs. $^{6}$He+$^{8}$Be) completely different optical model potentials
seem to be responsible for scattering in these two channels.  They
may reflect an adiabatic nucleon motion during the
$^{7}$Li($^{7}$Li,$^{6}$He)$^{8}$Be transfer reaction in which the
deep entrance channel potential is modified in shape and depth to
become the shallow one in the exit channel.

\par
     In Fig. \ref{fig:F07} the most shallow one from  equivalent $^{7}$Li+$^{7}$Li
OM potentials was used in the exit channel ("3 MeV" family), but
different potentials were applied in the entrance channel. Again, as
in Fig. \ref{fig:F05} the shallow potentials in the entrance channel
overestimate the cross sections. The potentials deeper than 60 MeV
reproduce the experimental angular distributions quite well.
Moreover, it can be seen that the agreement of {\it prior} and {\it
post} representations is better assured by deep entrance channel
potentials than by shallow ones (with the exception of identical
entrance and exit channel potentials; see above).  In summary we
conclude that the $^{7}$Li($^{7}$Li,$^{6}$He)$^{8}$Be reaction
favors rather deep OM potentials in the entrance and shallow ones
for the exit channel.
\par
\begin{figure}[ht!]
  \begin{center}
  \includegraphics[width=0.5\textwidth]{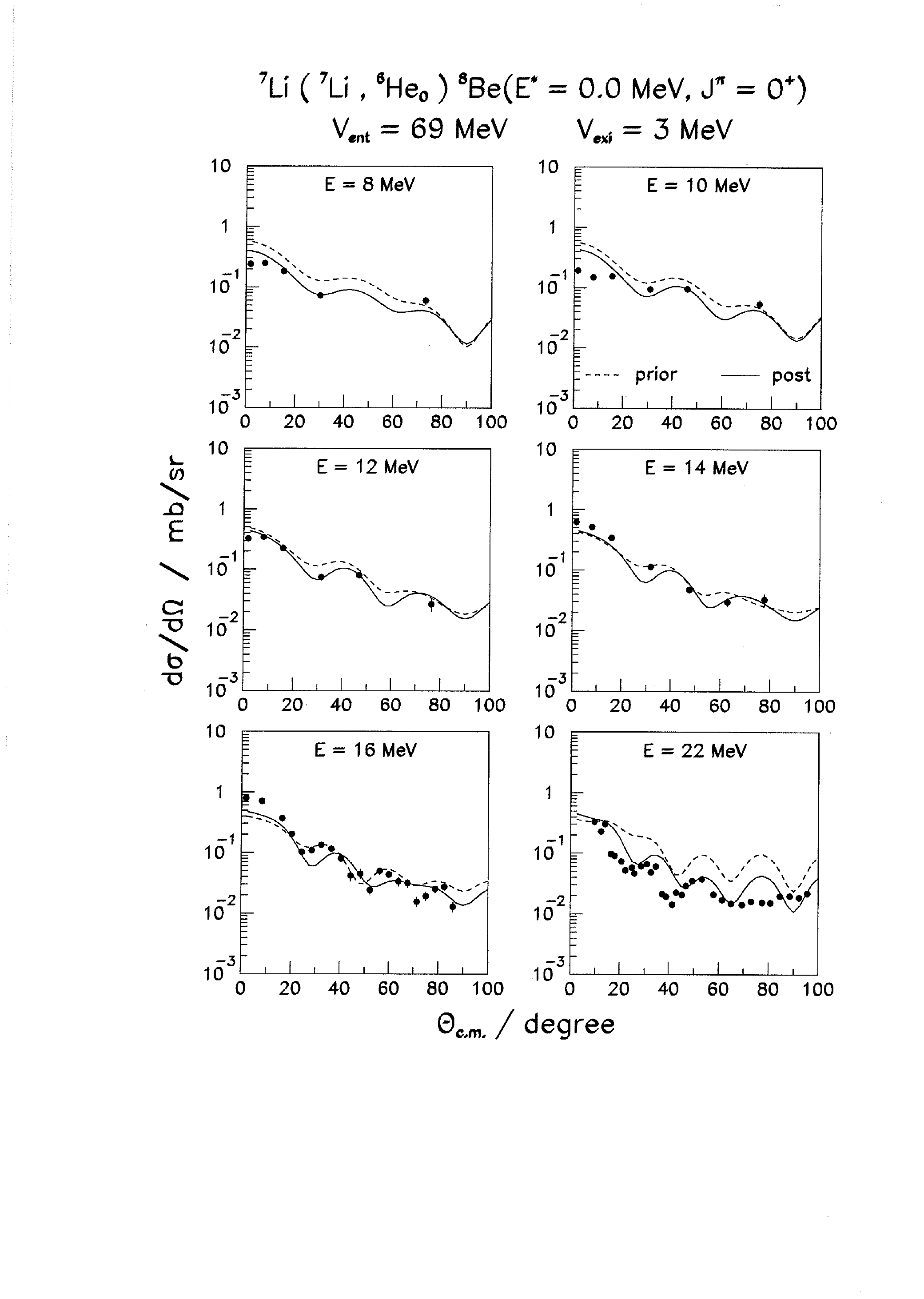}\\
  \caption{ Experimental angular distributions of the
         $^{7}$Li($^{7}$Li,$^{6}$He)$^{8}$Be$_{g.s.}$ reaction for
         several bombarding
         energies (full dots, present work) together with
         results of calculations performed with selected pairs
         of OM potentials: deep potential (family \#4, i.e.
         "69 MeV" of \cite{BAC93}) in the entrance channel and a
         very shallow one (family \#1, i.e. "3 MeV" of \cite{BAC93})
         in the exit channel.  Full lines correspond to
         {\it post}, dashed lines to {\it prior} representations.
         The data at E$_{lab}$=22 MeV were taken from ref. \cite{BOC88}.}\label{fig:F08}
  \end{center}
\end{figure}

     In Fig. \ref{fig:F08} we present results of the calculations which
were performed for several bombarding energies between 8 MeV (lab.)
and 22 MeV with this selected pair of OM potentials. The
reproduction of the experimental angular distributions may be judged
as very good in particular in view of the fact that the calculations
were carried out without any free parameters and that cluster
spectroscopic factors are known only with some (model dependent)
accuracy.

%----------------------------------------------------------------------------
%----------------------------------------------------------------------

\begin{figure}[ht!]
  \begin{center}
  \includegraphics[width=0.5\textwidth]{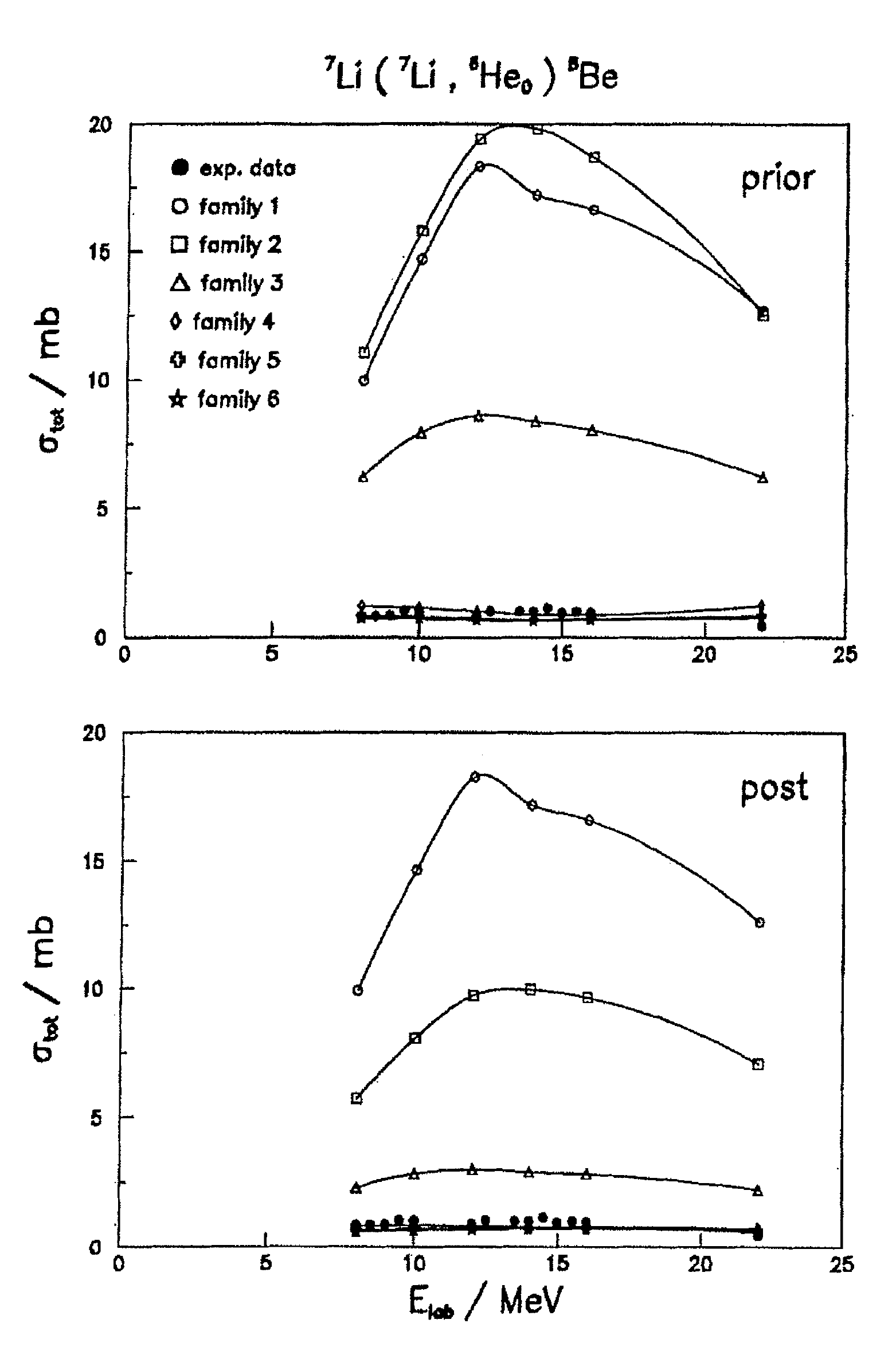}\\
  \caption{\label{fig:F09} Experimental angle integrated cross sections for
the $^{7}$Li($^{7}$Li,$^{6}$He)$^{8}$Be$_{g.s.}(0^{+})$ reaction as
a function of projectile energy (solid dots) and results of DWBA
calculations performed in  \emph{prior} (upper part of the figure)
and in \emph{post} representations (lower part).  Different lines
correspond to the use of different OM potentials (ref. \cite{BAC93})
for the entrance channel and the same OM potential (family \#1, i.e.
"3 MeV" of \cite{BAC93}) for the exit channel.}
  \end{center}
\end{figure}

This good agreement is also seen in the energy dependence of angle
integrated cross sections which are shown in Fig. \ref{fig:F09}. The
dots represent the experimental data of the present work, lines show
results of DWBA calculations for {\it prior} (upper part of the
figure) and {\it post} representation (lower part of the figure)

%----------------------------------------------------------------------------
%----------------------------------------------------------------------

calculated with different OM potentials for the entrance
$^{7}$Li+$^{7}$Li channel and with the same, shallow potential ("3
MeV" family) in the exit channel.

%----------------------------------------------------------------------------
\begin{figure}[ht!]
  \begin{center}
  \includegraphics[width=0.5\textwidth]{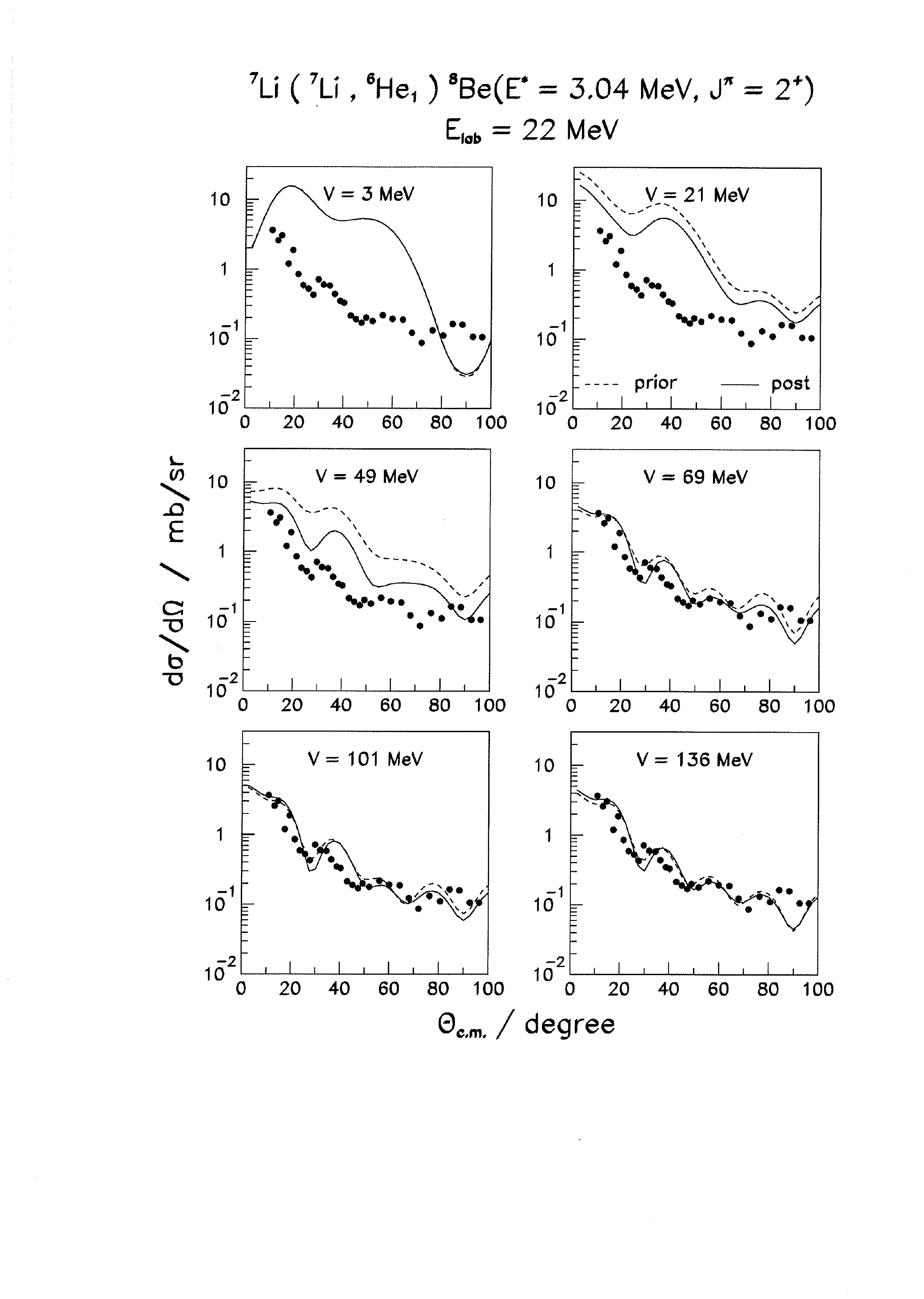}\\
  \caption{\label{fig:F10} Experimental angular distribution for the
         $^{7}$Li($^{7}$Li,$^{6}$He)$^{8}$Be$_{3.04}$ reaction
         at 22 MeV (lab.) \cite{BOC88}
         (full dots) and results of DWBA calculations performed
         with a very shallow (family \#1, i.e. "3 MeV" of ref. \cite{BAC93})
         $^{6}$He-$^{8}$Be OM potential and with various
         $^{7}$Li-$^{7}$Li OM potentials taken also from ref. \cite{BAC93}.}
  \end{center}
\end{figure}
%----------------------------------------------------------------------

%
\par
     The analysis of the reaction leading to the first excited
state of $^{8}$Be may yield an additional test of the proton
transfer mechanism.  In our experiment we were not able to evaluate
properly the differential cross sections from the experimental
spectra but there are data available of Bochkarev et al.
\cite{BOC88} measured for the
$^{7}$Li($^{7}$Li,$^{6}$He)$^{8}$Be$_{3.04}$ reaction at 22 MeV
(lab.).  Calculations were performed along the lines which yielded
the results shown in Figs. 5 - 8.  A very similar picture arises.
Therefore we present in Fig. \ref{fig:F10} only results of
calculations obtained with very shallow $^{6}$He-$^{8}$Be OM
potential ("3 MeV" family) and with various $^{7}$Li-$^{7}$Li OM
potentials.  Again only rather deep entrance channel potentials can
well describe the experimental data and the quality of this
description is quite satisfactory.
\par
     From these results we conclude the direct proton
transfer to dominate in $^{7}$Li($^{7}$Li,$^{6}$He)$^{8}$Be
reactions and the distorted wave Born approximation to be able to
properly describe experimental cross sections in the studied range
of energies. The somewhat poorer reproduction of 22 MeV angular
distributions may indicate a need for some energy dependence of the
optical model potentials (which were fitted to elastic scattering
data independently of energy in a restricted range of $^{7}$Li
projectile energies i.e. 8 - 17 MeV in the lab. system
\cite{BAC93}).

\smallskip

\subsection {\label{sec:alpha} The alpha particle transfer reaction -
$^{7}$Li($^{7}$Li,t)$^{11}$B }

\smallskip
\par
     This reaction seems to be less suited for DWBA in
comparison with the proton transfer reaction since the rearrangement
of nucleons during alpha particle transfer is more drastic than in
proton transfer.  Thus it is not clear whether it is appropriate to
treat alpha particle transfer in such a light nuclear system as a
perturbation.   Furthermore, cross sections of alpha particle
transfer are typically larger by an order of magnitude than those
for proton transfer as can be seen in the Fig.\ref{fig:F03}.  They
represent approximately 10\% of the elastic scattering cross
section, a fact which may disqualify the perturbation approach.
\par
     The DWBA analysis of the alpha particle transfer was
performed along the same lines delineated for proton transfer.
Again, no free parameters were allowed for optical model potentials,
spectroscopic amplitudes and transition potentials.  It was found
that results of the calculations are only weakly sensitive to the
exit channel t+$^{11}$B optical model potential.  Thus, in the
systematic calculations only one OM potential, taken from the
literature (ref. \cite{HAR80}), was used for generating distorted
waves in the exit channel.

%----------------------------------------------------------------------------
\begin{figure}[ht!]
  \begin{center}
  \includegraphics[width=0.5\textwidth]{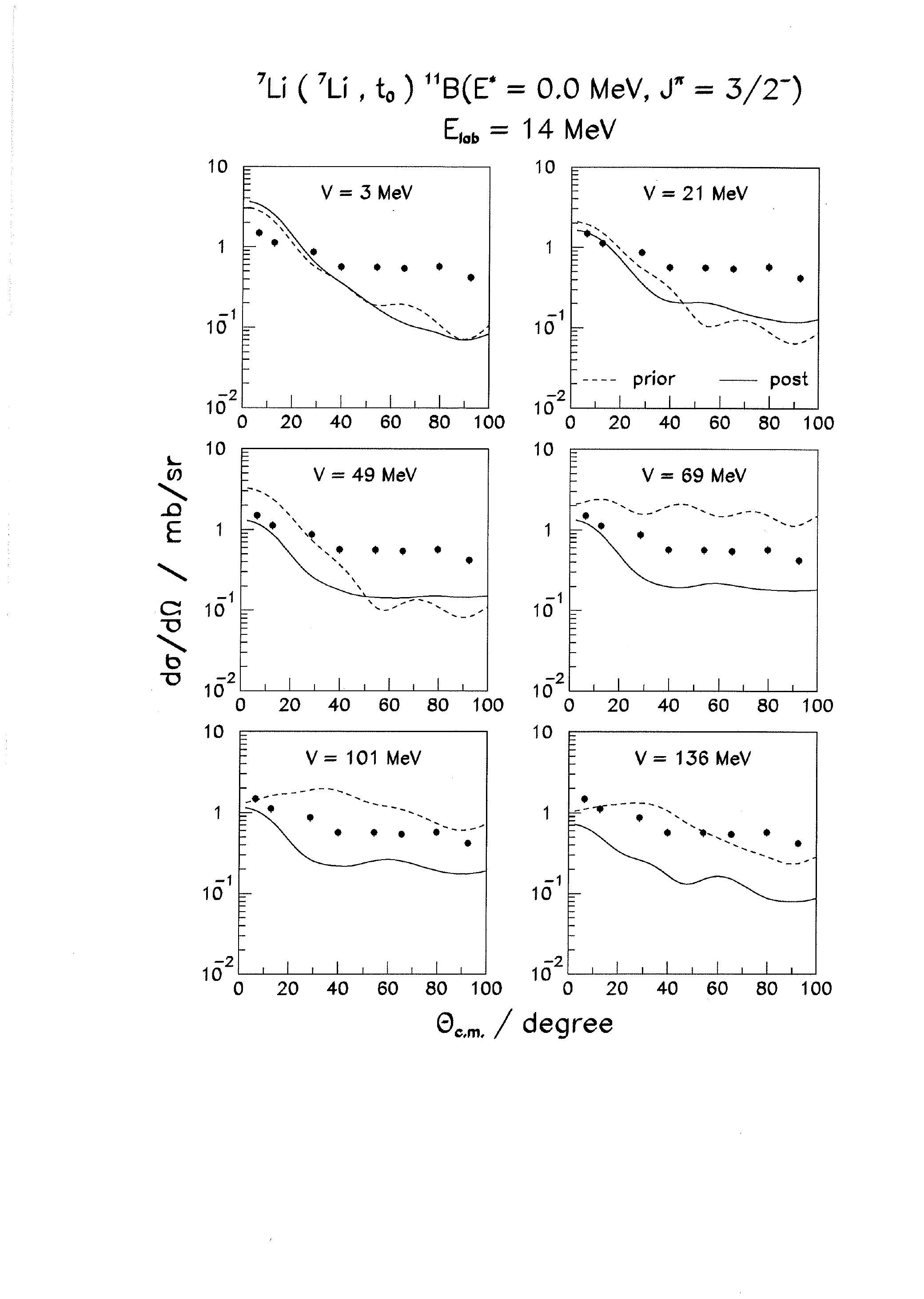}\\
  \caption{\label{fig:F11} Experimental angular distribution for the
  $^{7}$Li($^{7}$Li,t)$^{11}$B$_{g.s.}$(3/2$^{-}$) reaction
  at 14 MeV projectile energy (solid dots), and the results of DWBA
  calculations performed in {\it prior} (dashed lines) and in
  {\it post}  representation (full lines).  Different
  frames in the figure correspond to calculations
  performed with the same $^{11}$B-t OM potential (ref.
  \cite{HAR80}) but with different $^{7}$Li-$^{7}$Li OM potentials
  of ref. \cite{BAC93}.  The depth of the real part of
  these potentials is given in the corresponding
  frames.}
  \end{center}
\end{figure}
%----------------------------------------------------------------------

However, it was found that optical model potentials of the entrance
channel which equally well reproduce elastic scattering produce
quite different results when applied in the DWBA.  Hence, for
comparison, the calculations were performed with the same six OM
potentials of the $^{7}$Li+$^{7}$Li channel which were already used
in the analysis of the proton transfer reaction.

%----------------------------------------------------------------------------
\begin{figure}[ht!]
  \begin{center}
  \includegraphics[width=0.5\textwidth]{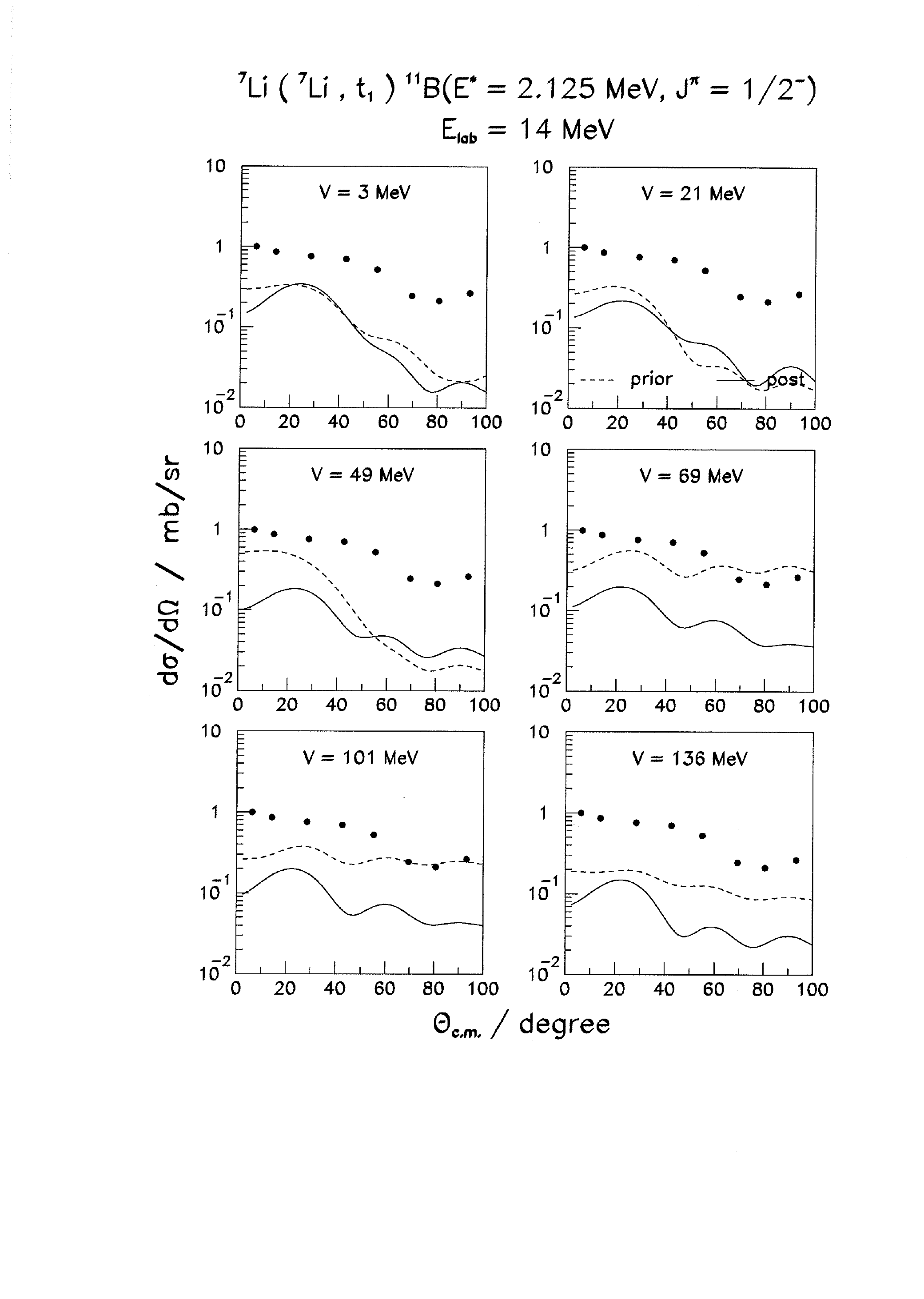}
  \caption{\label{fig:F12} Same as Fig. \ref{fig:F11}, but for the
         $^{7}$Li($^{7}$Li,t)$^{11}$B$_{2.13}$(1/2$^{-}$) reaction
         at E$_{lab}$=14 MeV.}
  \end{center}
\end{figure}
%----------------------------------------------------------------------

%
\par
     One may conjecture by inspecting Fig. \ref{fig:F03} that
contributions of mechanisms which are characterized by a strong
energy dependence of their cross section e.g. statistical
fluctuations and/or excitation of individual resonances, are present
in this reaction but least important at the highest energy.
Therefore we need to compare both angular distributions and
excitation functions as given by DWBA with experimental data.

\par
\begin{figure}[t!]
  \begin{center}
  \includegraphics[width=0.48\textwidth]{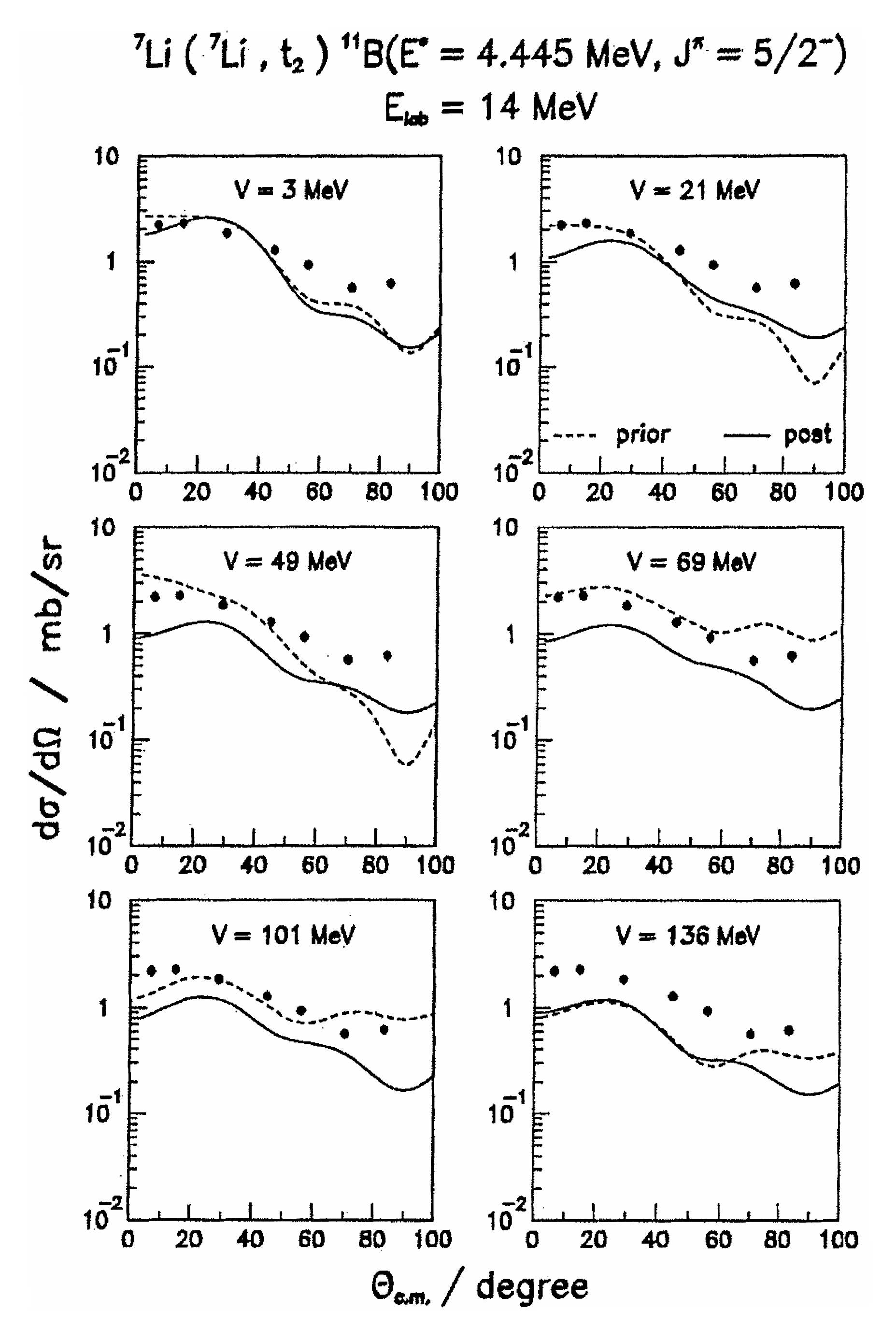}\\
  \caption{\label{fig:F13}  Same as Fig. \ref{fig:F11}, but for the
         $^{7}$Li($^{7}$Li,t)$^{11}$B$_{4.45}$(5/2$^{-}$) reaction
         at E$_{lab}$=14 MeV.
}
  \end{center}
\end{figure}
\begin{figure}[ht!]
  \begin{center}
  \includegraphics[width=0.5\textwidth]{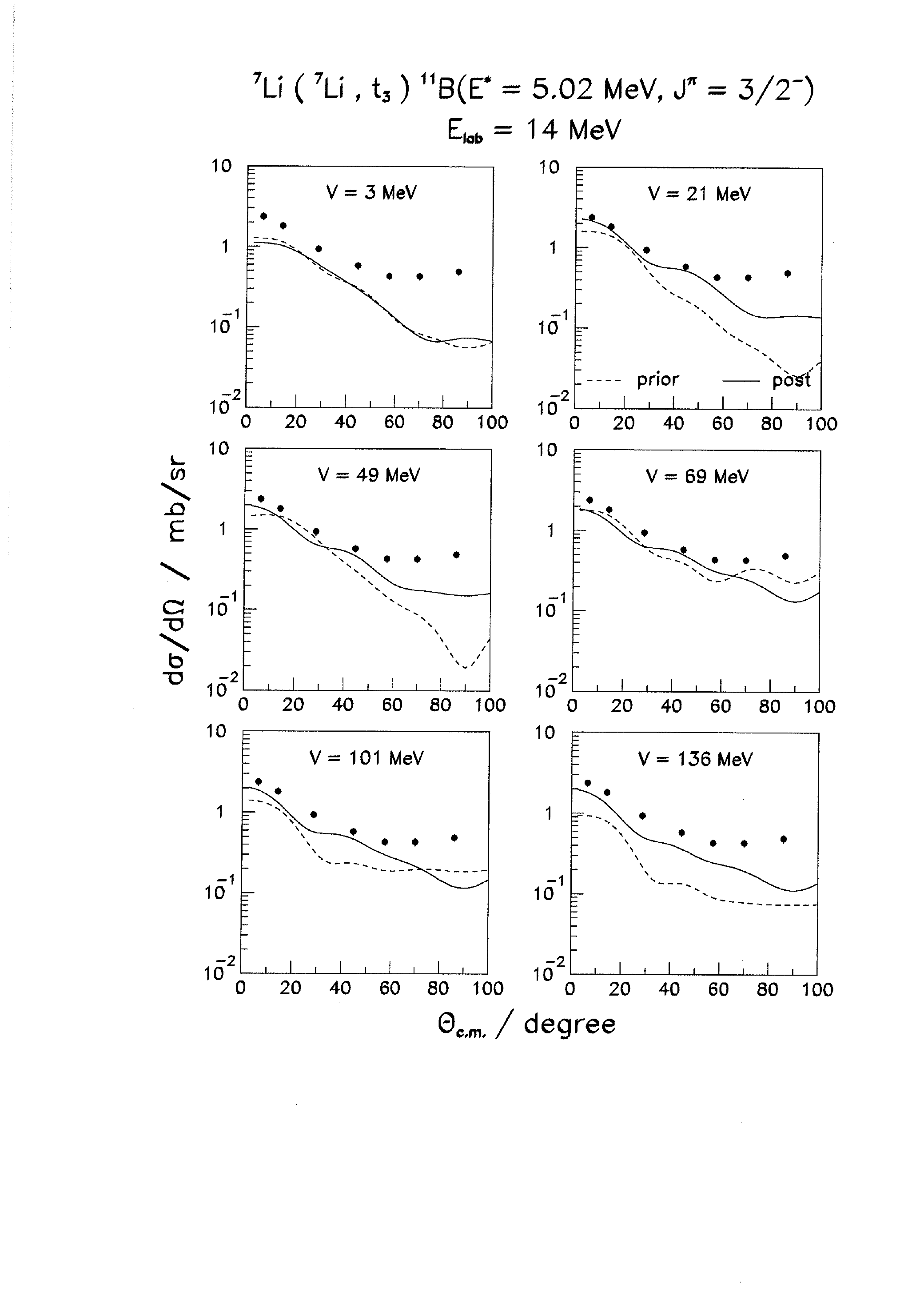}\\
  \caption{\label{fig:F14} Same as Fig. \ref{fig:F11}, but for the
         $^{7}$Li($^{7}$Li,t)$^{11}$B$_{5.02}$(3/2$^{-}$) reaction
         at E$_{lab}$=14 MeV.}
  \end{center}
\end{figure}

     Results of the calculations performed at E$_{lab}$=14 MeV are
presented in Fig. \ref{fig:F11} for the ground state transfer
$^{7}$Li($^{7}$Li,t)$^{11}$B$_{g.s.}$ and in Figs. \ref{fig:F12},
\ref{fig:F13} and \ref{fig:F14} for the the transfer to the first
(2.13 MeV; 1/2$^{-}$), second (4.45 MeV; 5/2$^{-}$), and third (5.02
MeV; 3/2$^{-}$) excited state, respectively.

%----------------------------------------------------------------------------
%----------------------------------------------------------------------

The description of the experimental angular distributions by DWBA is
acceptable but significantly poorer than for proton transfer.
Furthermore, the prior - post equivalence is not well established
(especially for deep OM potentials) pointing at the limits of
accuracy of the DWBA approach for alpha particle transfer in the
$^{7}$Li+$^{7}$Li system.

%----------------------------------------------------------------------------
%----------------------------------------------------------------------

\par
\begin{figure}[ht!]
  \begin{center}
  \includegraphics[width=0.48\textwidth]{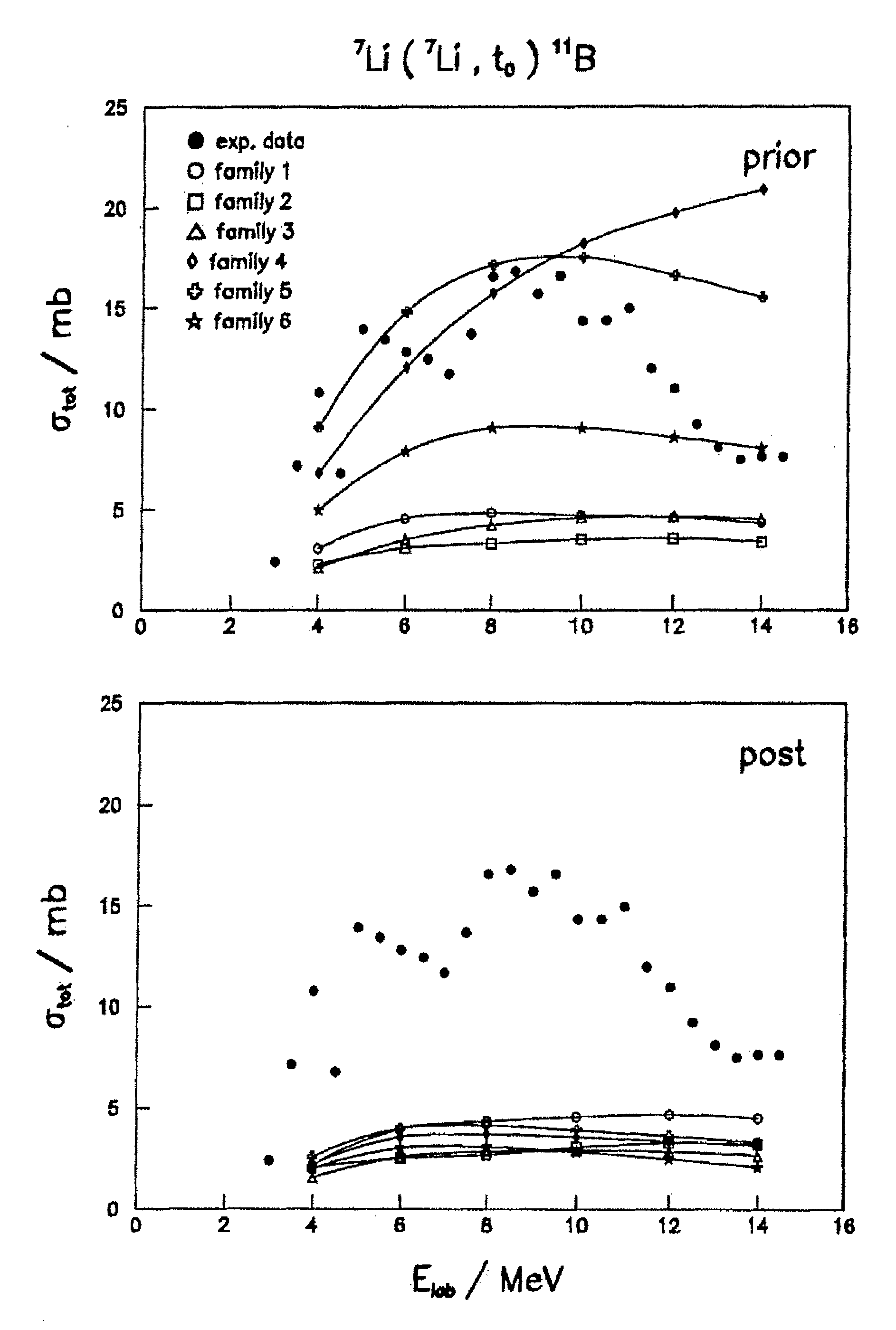}\\
  \caption{\label{fig:F15} Experimental angle integrated cross section for the
         $^{7}$Li($^{7}$Li,t)$^{11}$B$_{g.s.}$(3/2$^{-}$) reaction
         as a function of projectile energy (solid dots) and results of DWBA
         calculations performed in {\it prior} (upper part of the figure)
         and in {\it post} representations (lower part).
         Different lines correspond to the
         use of different OM potentials for the entrance
         channel from ref. \cite{BAC93} and the same OM potential
         (ref. \cite{HAR80}) for the exit channel.}
  \end{center}
\end{figure}
\begin{figure}[t!]
  \begin{center}
  \includegraphics[width=0.48\textwidth]{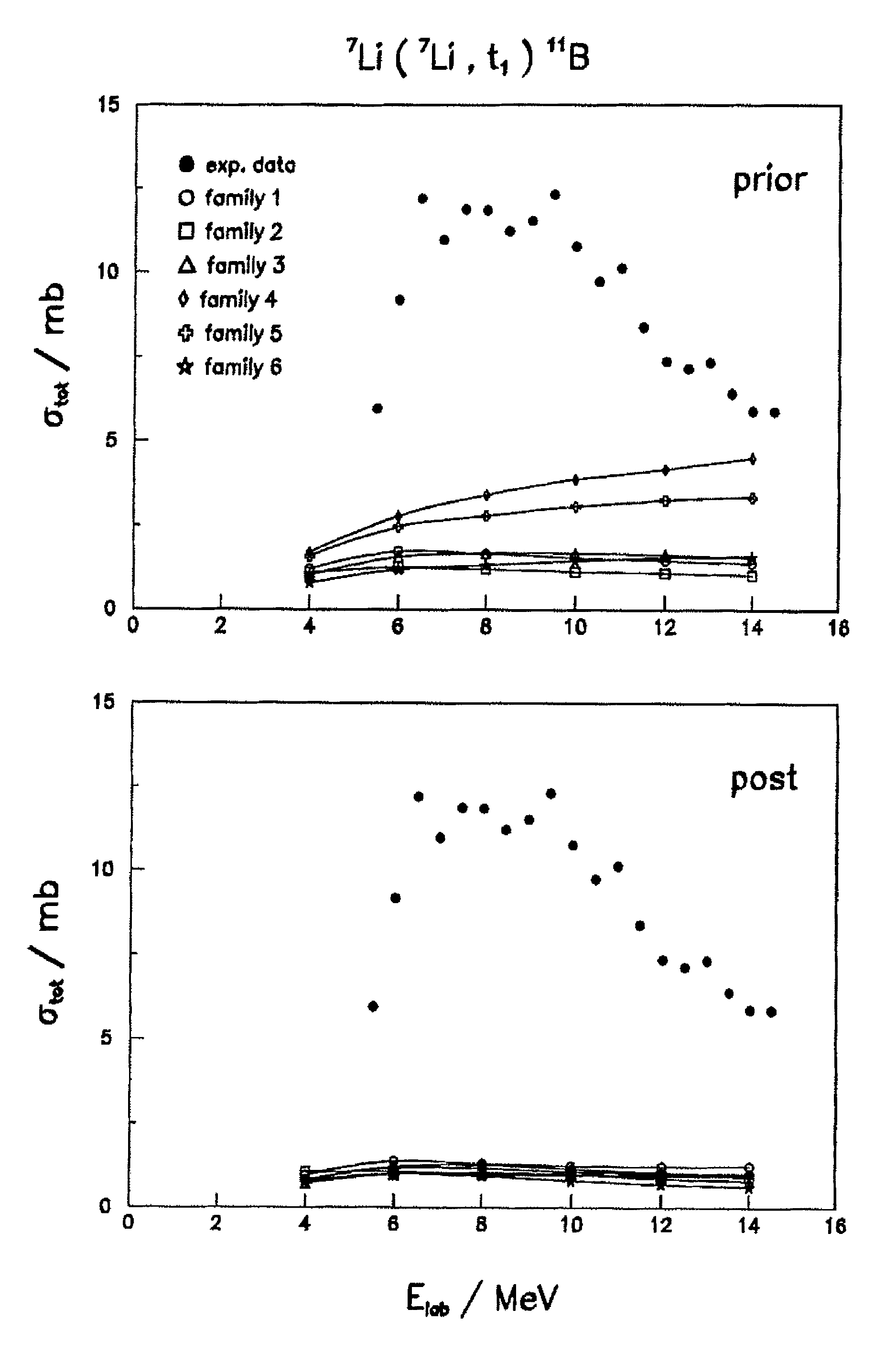}\\
  \caption{\label{fig:F16} Same as Fig. \ref{fig:F15}, but for the
         $^{7}$Li($^{7}$Li,t)$^{11}$B$_{2.13}(1/2^{-}$) reaction.}
  \end{center}
\end{figure}
\begin{figure}[t!]
  \begin{center}
  \includegraphics[width=0.48\textwidth]{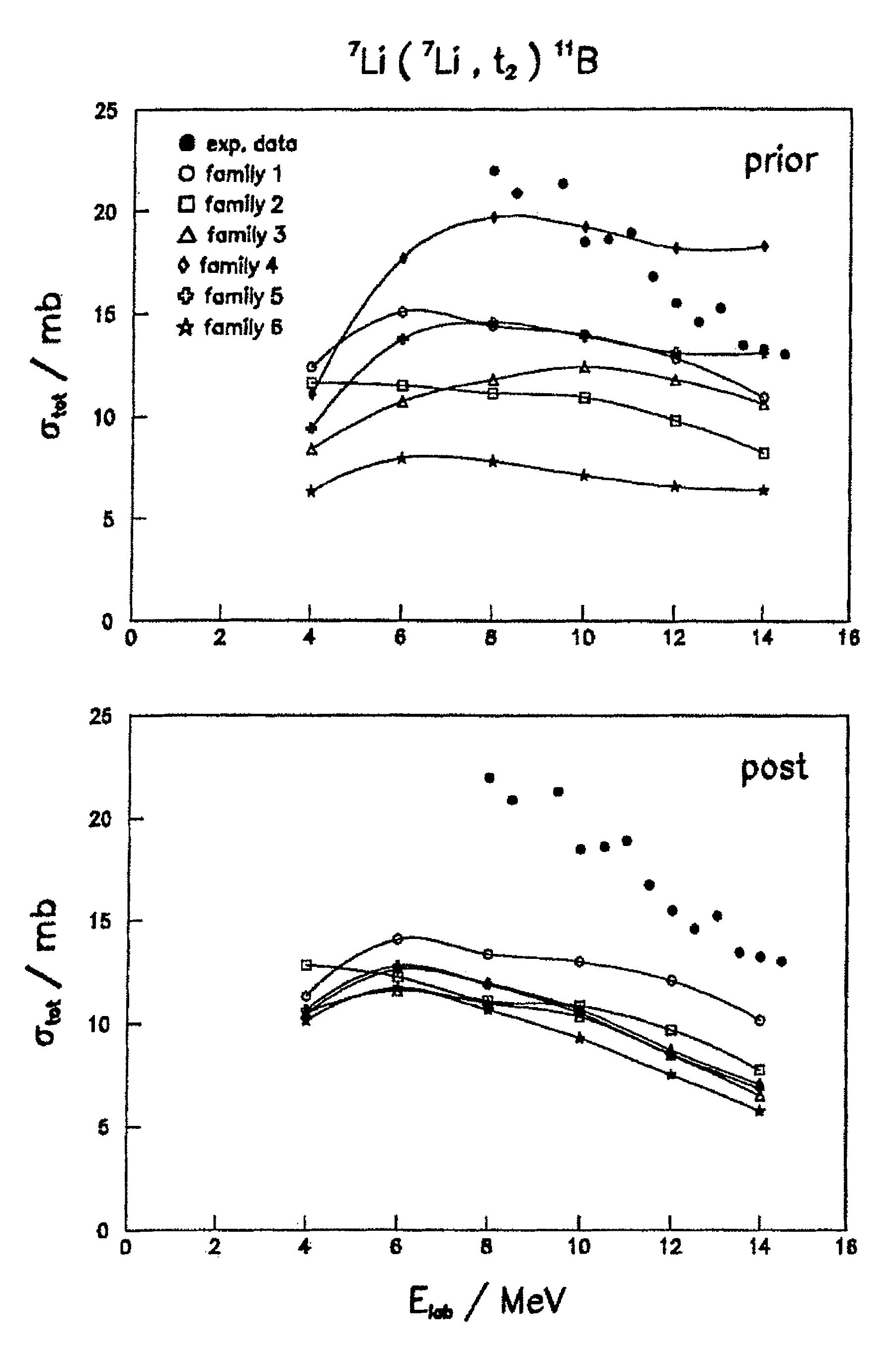}\\
  \caption{\label{fig:F17} Same as Fig. \ref{fig:F15}, but for the
         $^{7}$Li($^{7}$Li,t)$^{11}$B$_{4.45}(5/2^{-}$) reaction.}
  \end{center}
\end{figure}
\begin{figure}[t!]
  \begin{center}
  \includegraphics[width=0.48\textwidth]{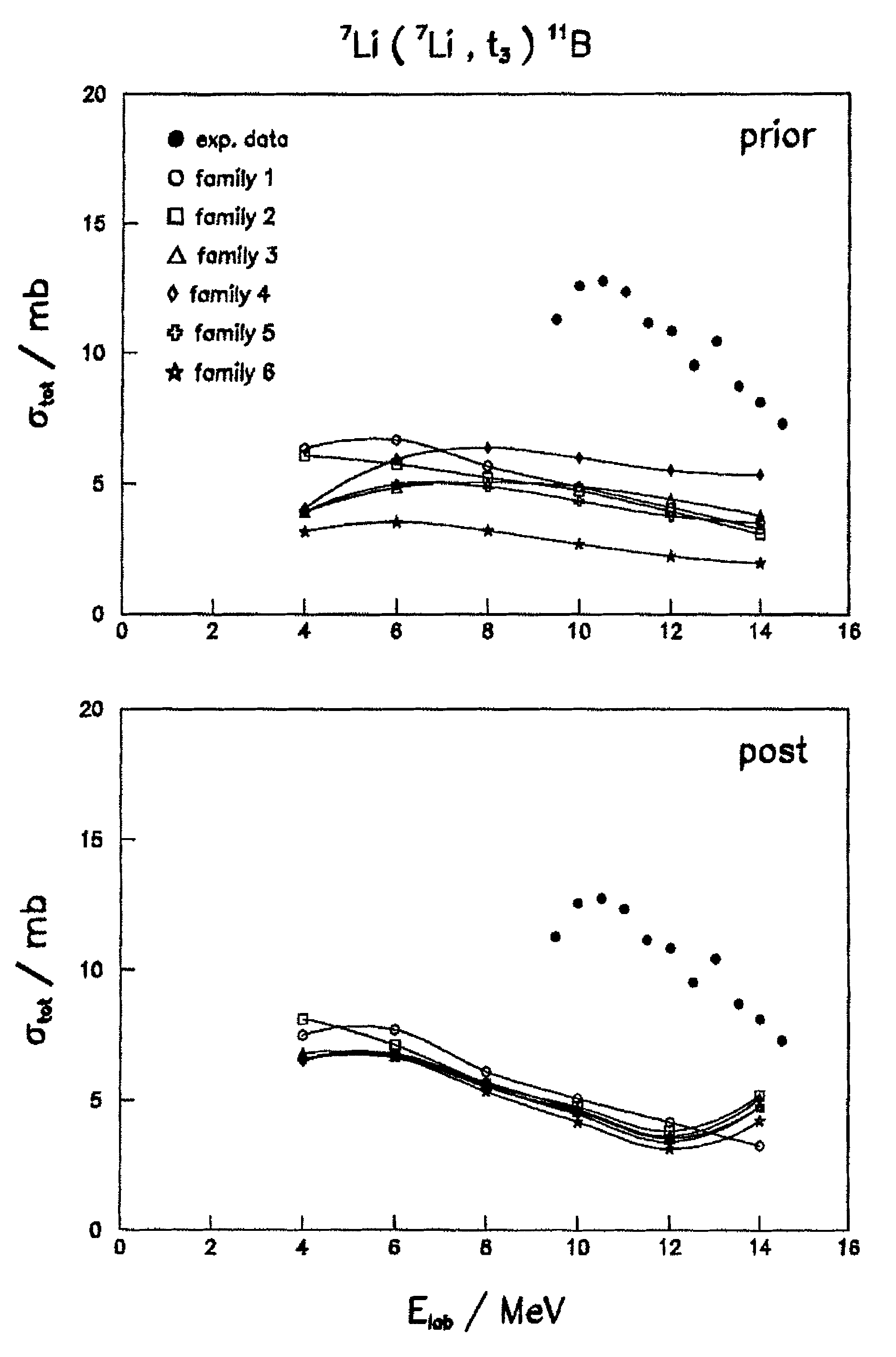}\\
  \caption{\label{fig:F18} Same as Fig. \ref{fig:F15}, but for the
         $^{7}$Li($^{7}$Li,t)$^{11}$B$_{5.02}(3/2^{-})$ reaction.}
  \end{center}
\end{figure}

     In Figs. \ref{fig:F15} - \ref{fig:F18} the angle integrated cross sections are
depicted versus beam energy.  Black dots represent experimental
cross sections and the lines correspond to DWBA calculations
performed with different $^{7}$Li-$^{7}$Li OM potentials (cf. Table
\ref{tab:1}) in {\it prior} (upper part of the figure) and {\it
post} (lower part of the figure) representations.  It may be
concluded that results obtained in  {\it post} representation are in
general more consistent than those in the {\it prior} one, i.e. the
angle integrated cross sections for all OM potentials yield similar
values.

%----------------------------------------------------------------------------
%----------------------------------------------------------------------

Moreover, the "post - prior" equivalence which was reasonably well
fulfilled at E$_{lab}$=14 MeV for shallow potentials remains to be
fulfilled for these potentials in the whole energy range.  Thus, one
has either to use the {\it post} representation or to choose the
shallow potentials for calculations in the {\it prior}
representation.  In these cases the shape of the energy dependence
of experimental cross sections is reasonably well reproduced by DWBA
calculations at least for higher energies. The size of the cross
section, however, is predicted too small, clearly showing the
presence of other reaction mechanisms.  DWBA gives an almost
constant value of the angle integrated cross section for the ground
state transition while the experimental one varies very strongly
with energy and has a maximum about E$_{lab}$= 8 MeV.

%----------------------------------------------------------------------------
%----------------------------------------------------------------------

  Resonances
are likely to contribute to the triton channel in the low energy
region. They exhaust the major part of the experimental cross
section.  Only at the highest energy of E$_{lab}$=14 MeV direct
reactions become important with contributions of approximately 50\%
for the ground state transition, and 20 \% , 50\% and 60 \% for the
transitions leading to the first, the second and the third excited
states of $^{11}$B, respectively.

%----------------------------------------------------------------------------
%----------------------------------------------------------------------

%----------------------------------------------------------------------------
%----------------------------------------------------------------------

\smallskip

\subsection{\label{sec:triton} The triton transfer reaction
- $^{7}$Li($^{7}$Li,$^{4}$He)$^{10}$Be}

\smallskip
\par
     It is apparent from examining Figs. 3 and 4 that a strong
variation of the experimental angle integrated cross sections
versus energy is present for the $^{7}$Li($^{7}$Li,$^{4}$He) reactions.
Therefore this channel is the least suited among all reactions
under investigation for the application of the direct reaction
formalism.  Led by results for the alpha particle transfer
reactions we can expect DWBA to be appropriate at the
highest energy (E$_{lab}$=14 MeV).  DWBA calculations were performed
at this energy for transitions to both the ground state of
$^{10}$Be and the first excited state $^{10}$Be$_{3.37}(2^{+}$).

\par
     It turned out that in these cases the results of DWBA
calculations depend only weakly on the parameters of the exit
channel ($^{4}$He+$^{10}$Be) optical model potential.  Hence, only one
OM potential was applied for evaluation of distorted waves
in the exit channel \cite{ENG77}.  All six OM potentials
(cf. Table \ref{tab:1}) used previously in the analysis of the
proton
and the alpha particle transfer were exploited for generation
of distorted waves in the entrance channel.

%----------------------------------------------------------------------------
\begin{figure}[t!]
  \begin{center}
  \includegraphics[width=0.5\textwidth]{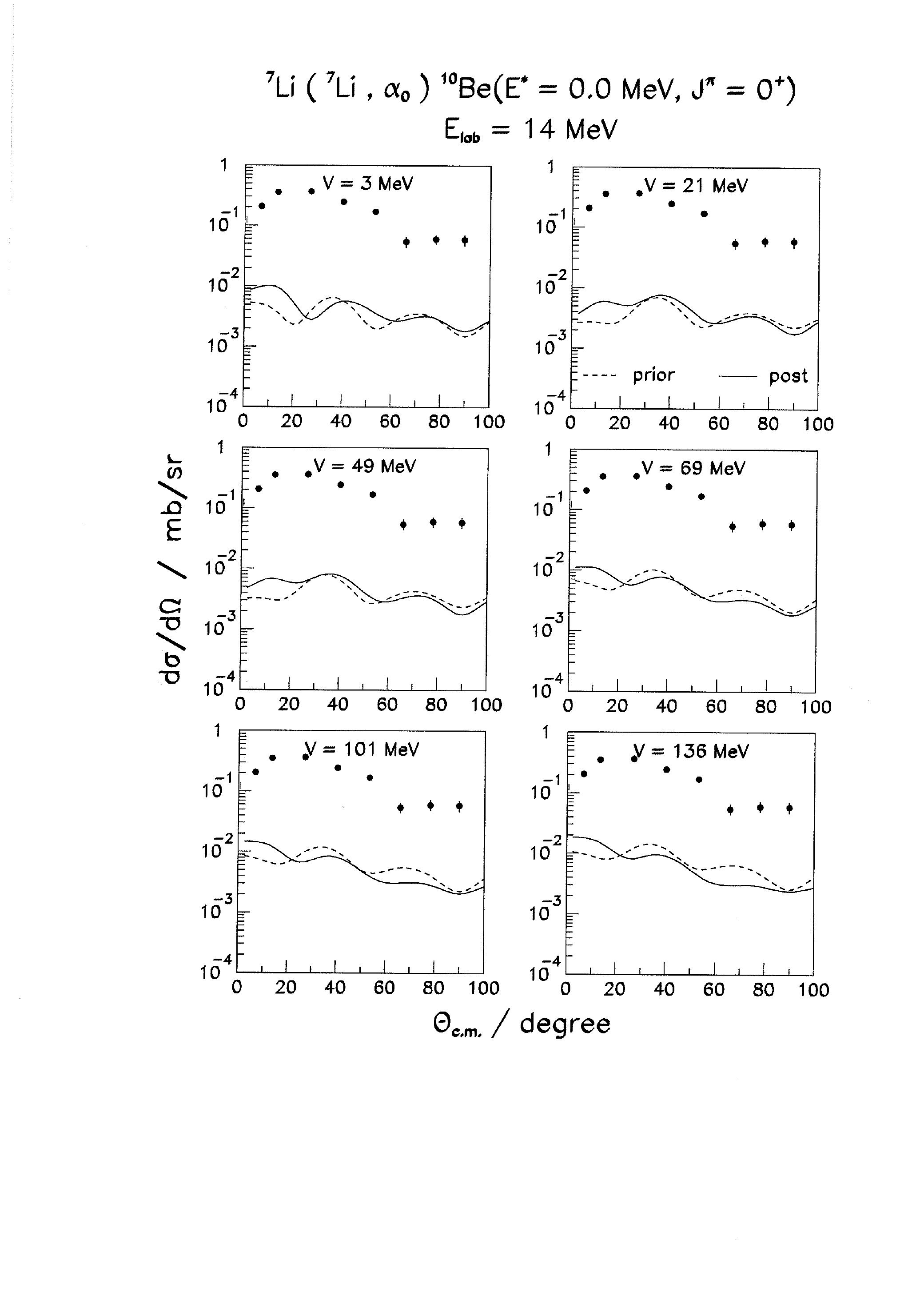}\\
  \caption{\label{fig:F19} Experimental angular distribution for the
         $^{7}$Li($^{7}$Li,$^{4}$He)$^{10}$Be$_{g.s.}(0^{+}$)
         reaction at E$_{lab}$=14 MeV
         (solid dots) and the results of DWBA
         calculations performed in {\it prior} (dashed lines)
         and in {\it post} representation (full lines).
         Different frames in the figure correspond to
         calculations performed with the same $^{10}$Be-$^{4}$He OM
         potential (ref. \cite{ENG77}) but with different $^{7}$Li-$^{7}$Li
         OM potentials from ref. \cite{BAC93}.
         The depth of the real part of these potentials is
         given in the corresponding frames.}
  \end{center}
\end{figure}
%----------------------------------------------------------------------

%----------------------------------------------------------------------------
\begin{figure}[t!]
  \begin{center}
  \includegraphics[width=0.5\textwidth]{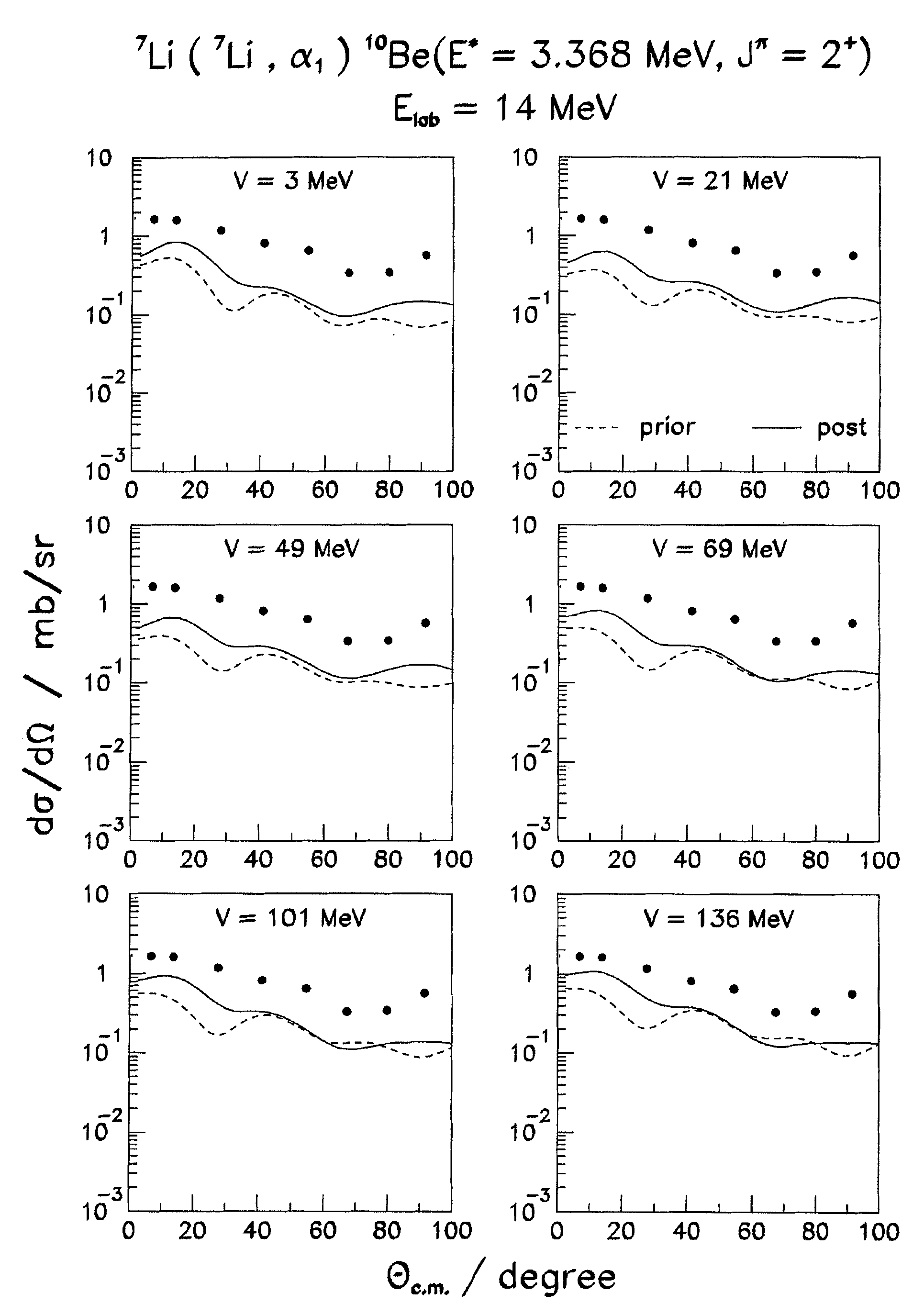}\\
  \caption{\label{fig:F20} Same as Fig. \ref{fig:F19}, but for the the
         $^{7}$Li($^{7}$Li,$^{4}$He)$^{10}$Be$_{3.37}(2^{+})$ reaction
         at E$_{lab}$=14 MeV.
}
  \end{center}
\end{figure}
%----------------------------------------------------------------------

\par
     The experimental angular distributions (full dots) are
shown in Figs. 19 and 20 together with results of DWBA
calculations performed using the {\it prior} (dashed lines) and
the {\it post} representation (full lines) for
$^{7}$Li($^{7}$Li,$^{4}$He)$^{10}$Be$_{g.s.}$(0$^{+}$) and
$^{7}$Li($^{7}$Li,$^{4}$He)$^{10}$Be$_{3.37}$(2$^{+}$)
reactions, respectively.  The calculations in both {\it prior} and
{\it post} representation result in angular distributions similar
in shape as well as in magnitude.  DWBA angular distributions
are in either case smooth with small oscillations
reproducing qualitatively the shape of the experimental
angular distributions.  The magnitude of the theoretical
cross sections is, however, in either case smaller than that
for the experimental data.  In case of the ground state
transition the theoretical cross section is smaller by a factor
40 - 50 than the experimental cross section and for
alpha transfer leading to the first excited state of $^{10}$Be the
DWBA cross section exhausts approximately 40 - 50 \% of the
experimental data.

\par
     To estimate the average contribution of a direct mechanism
to triton transfer the energy dependence of the angle
integrated DWBA cross section was calculated and compared
with the energy dependence of the experimental angle integrated
cross section.

%----------------------------------------------------------------------------
\begin{figure}[t!]
  \begin{center}
  \includegraphics[width=0.5\textwidth]{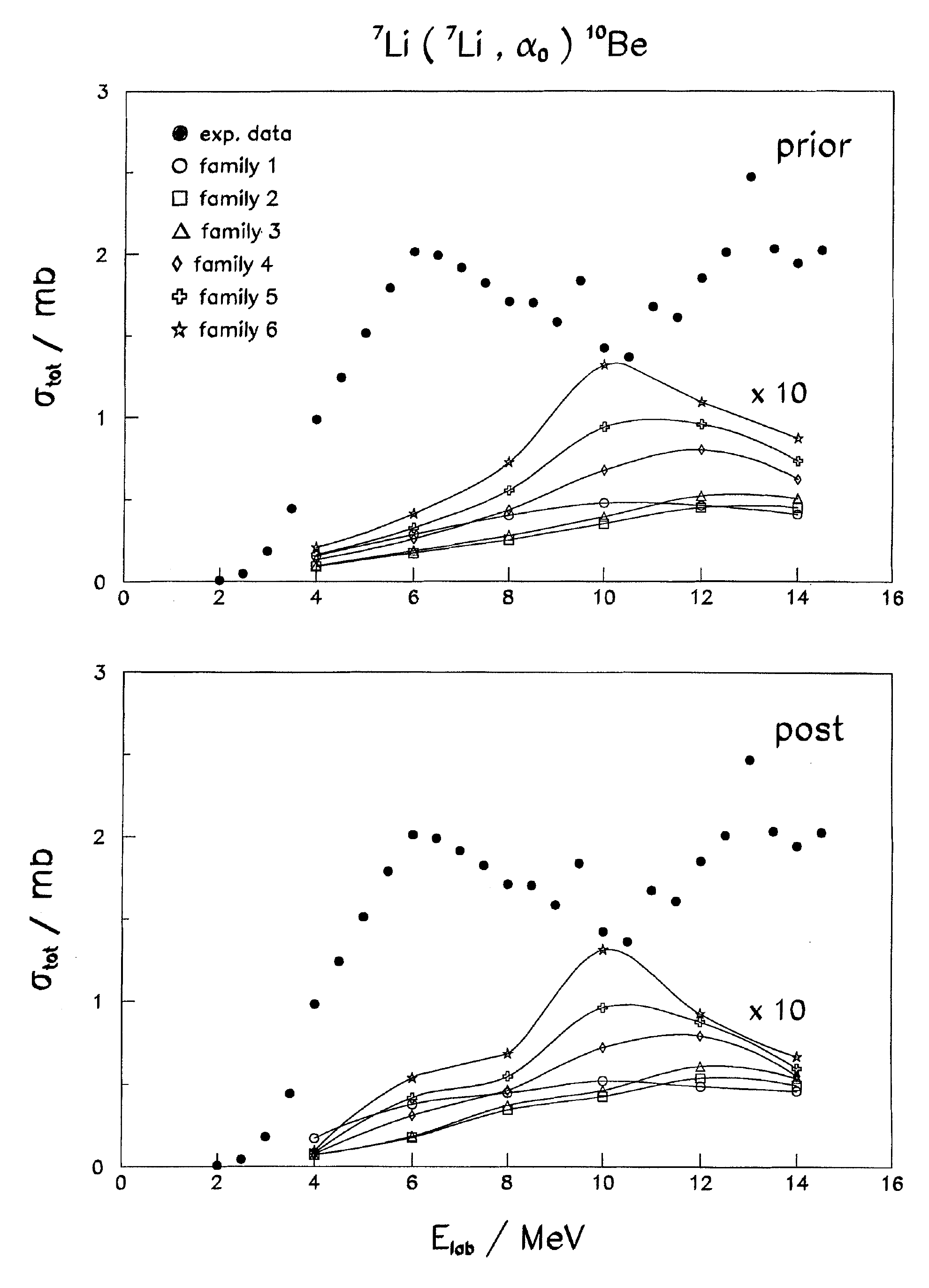}\\
  \caption{\label{fig:F21} Experimental, angle integrated cross sections for the
          $^{7}$Li($^{7}$Li,$^{4}$He)$^{10}$Be$_{g.s.}$(0$^{+}$) reaction
          as a function of
          projectile energy (solid dots), and results of DWBA
          calculations performed in {\it prior} (upper part of the figure)
          and in {\it post} representation (lower part).
          Different lines correspond to the
          use of different OM potentials (ref. \cite{BAC93}) for the
          entrance channel and the same OM potential (\cite{ENG77})
          for the exit channel.}
  \end{center}
\end{figure}
%----------------------------------------------------------------------

%----------------------------------------------------------------------------
\begin{figure}[t!]
  \begin{center}
  \includegraphics[width=0.455\textwidth]{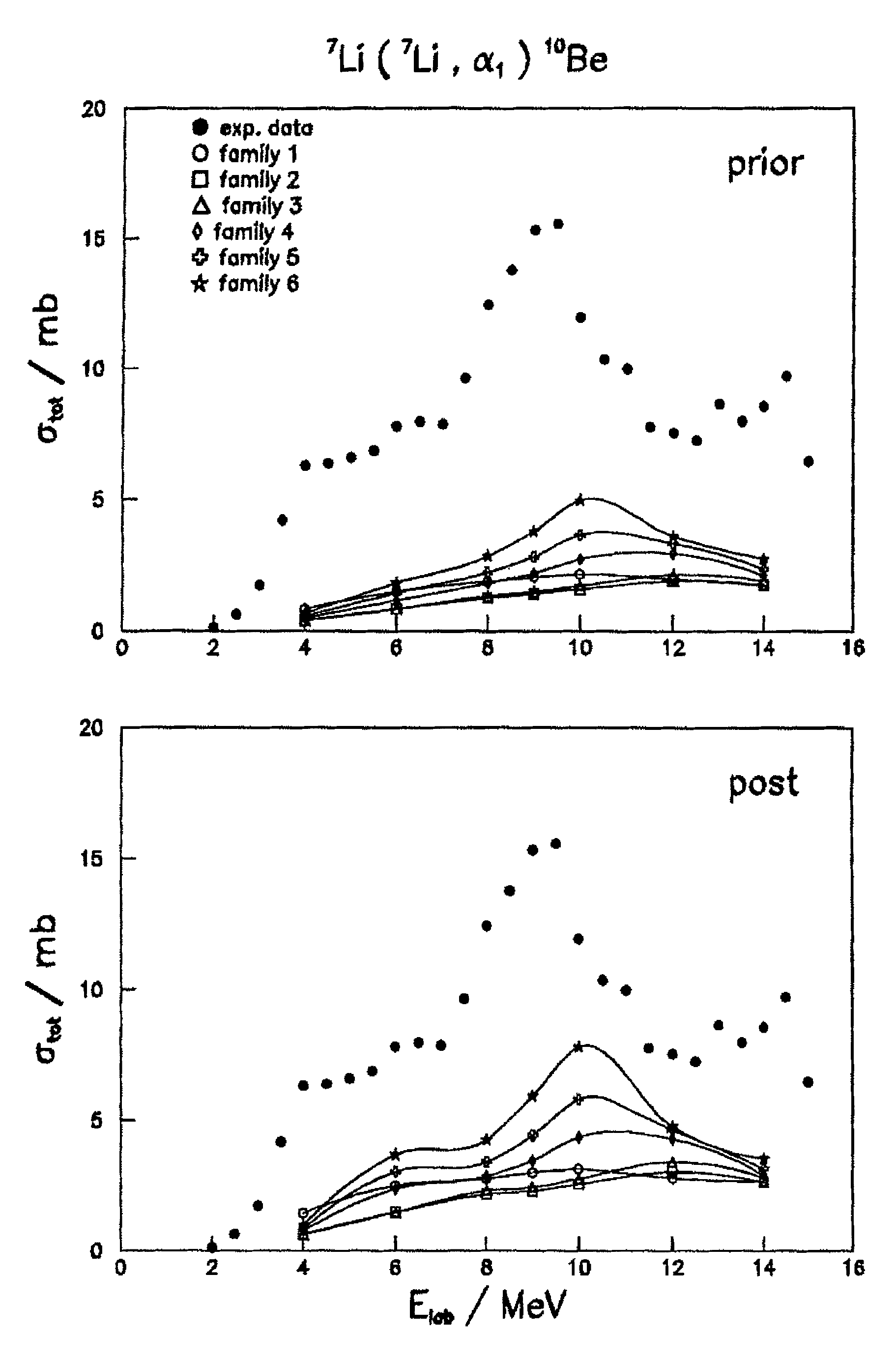}\\
  \caption{\label{fig:F22}
         Same as Fig. \ref{fig:F21}, but for the
         $^{7}$Li($^{7}$Li,$^{4}$He)$^{10}$Be$_{3.37}(2^{+}$) reaction.
}
  \end{center}
\end{figure}
%----------------------------------------------------------------------

This is illustrated by Figs. \ref{fig:F21} and \ref{fig:F22} for the
ground state transition and for the transition to the first excited
state of $^{10}$Be, respectively.  The theoretical cross section for
the ground state transition varies smoothly versus energy in the
investigated energy range.  It is, on average, smaller by a factor
20 - 40 than the experimental cross section. Note that the
theoretical cross section shown in Fig. \ref{fig:F21} is multiplied
by a factor of 10 for better representation. In contradistinction to
the DWBA predictions the experimental data vary rapidly with energy
(the experimental uncertainties are smaller than the dot size in the
figure).  This is also true for the transition to the first excited
state shown in Fig. \ref{fig:F22}. In this case, however, the
theoretical cross section establishes approximately 40 - 50 \% of
the experimental cross section.

\par
     It is interesting to note that the remaining parts of the
experimental cross sections which cannot be ascribed to a
direct reaction mechanism fulfill a simple $\it{" 2J+1 "}$
relationship for transitions to both the ground and the first
excited state.  Such a relationship is indicative for the
compound nucleus mechanism.  The strong energy dependence of
the cross sections seems to confirm this conjecture.

%\par
%     The large peak of the experimental angle integrated
%cross section present at an energy of E$_{lab}$= 9 MeV may be due to
%resonant formation of a highly excited compound system $^{14}$C.
%If this is the case the shape of the experimental
%angular distribution at this energy should be distinctively
%different from that predicted by DWBA.  To check this
%hypothesis we performed the appropriate DWBA calculations
%and the results are shown in the Fig. 23.

\section {\label{sec:summary} Summary and conclusions}
\hspace{1.cm}

In the present work experimental data are presented from a
measurement of angular distributions of $^7$Li($^7$Li,t)$^{11}$B,
$^7$Li($^7$Li,$^4$He)$^{10}$Be and $^7$Li($^7$Li,$^6$He)$^8$Be
reactions at several
energies between 8 and 16 MeV in the laboratory system.
Transitions to the ground states as well as to some low lying
excited states (three in the triton channel and one in the $^4$He
channel) were studied.  Already by inspection of these data we
are led to the conclusion that the ($^7$Li,$^6$He) reaction proceeds
predominantly as a direct process while various mechanisms may
contribute to ($^7$Li,t) and ($^7$Li,$\alpha$) reactions.  The latter
reaction seems to proceed predominantly through isolated
resonances of the $^{14}$C compound system.

\par
A DWBA analysis was performed for all channels, the
results confirm the qualitative conclusions derived from
inspection of the experimental data.  The ($^7$Li,$^6$He) reaction is
- within the accuracy of DWBA calculations - completely
described by direct proton transfer.  The other two reactions
proceed partially by direct mechanisms: in average 20 - 60 \% for
the $\alpha$-particle transfers ($^7$Li,t) and 40 - 50 \% for the
triton transfer ($^7$Li,$^4$He) reaction to the first excited state
of $^{10}$Be and only approximately 3 - 5 \% for transition to the
ground state.

\par
Estimations based on the Hauser-Feshbach model \cite{GOT89}
indicate a rather small contribution of compound nucleus reactions
(approximately 12-20\% for the triton channels and even less for
$\alpha$ particle and $^6$He channels).  Thus, processes different
than pure direct and pure compound nucleus mechanisms are present in
the investigated energy range.  The interference between direct and
compound nucleus reaction - amplitudes may lead to fluctuations of
the cross section which, however, are expected to be narrower
(typical width approx. 0.6 MeV) than the structures observed here.
Indeed, the presence of strong peaks in the excitation functions of
the angle integrated cross section for $\alpha_{1}$, correlated with
the structures visible in the excitation function measured at
forward angles suggests a contribution of isolated resonances
superimposed on the background from both direct and statistical
compound nucleus reactions.

\par
The good reproduction of both shape and magnitude of
the experimental angular distributions for the $^7$Li($^7$Li,$^6$He)$^8$Be
reaction by the DWBA calculations as well as the reproduction
of the energy dependence of the cross section indicates that
the methods of direct reaction theory can be successfully
applied for such a system of few nucleons.  The calculations
within both {\it prior} and {\it post} representations lead to
equivalent results and thus manifest the adequacy of the
DWBA approach to this reaction.  The use of different
optical model potentials in the DWBA calculations allowed
us to select from potentials which describe elastic and
inelastic scattering equally well those which are appropriate,
namely a rather deep (appr. 70 MeV) OM potential for the
entrance ($^7$Li+$^7$Li) and a very shallow potential (approx. 3
MeV) for the exit ($^6$He+$^8$Be) channel.  Only this combination of
OM potentials produces results which agree with the
experimental data of the $^7$Li($^7$Li,$^6$He)$^8$Be reaction.

\par
In spite of rather big cross sections for the $\alpha$
particle and triton channels (approximately 10\% of the
elastic cross section) and the relatively strong
rearrangement processes of the $^7$Li+$^7$Li system after triton
or $\alpha$ particle transfer the method of DWBA can be
successfully applied to evaluate the contribution of the
direct mechanism to the $^7$Li($^7$Li,$^4$He)$^{10}$Be and
$^7$Li($^7$Li,t)$^{11}$B
reactions.  Results of DWBA calculations turned out to be
insensitive to variation of the exit channel optical model
potentials in both cases but some caution is in order when
applying the {\it prior} or {\it post} representations together with
different OM potentials in the entrance channel.  In general
the {\it post} representation is superior and may be applied
without further selection of the optical model potentials.
The calculations in the {\it prior} representation lead to
equivalent results, as it is demanded by the DWBA formalism,
only if optical model potentials of the $^7$Li+$^7$Li system are
chosen which are rather shallow ($<$ 60 MeV ).  A selection
of $^7$Li+$^7$Li OM potentials on the basis of DWBA calculations
applied to triton and $\alpha$ particle transfers yields
results contrary to those obtained for the proton transfer.
This may indicate that either the proton transfer reaction
is sensitive to different parts of the OM potential than
the cluster transfers or the {\it prior} representation is not
well suited for these reactions e.g. due to poor cancellation
of "indirect transition" potentials.

\par
It should be emphasized that the good description of the
experimental data by DWBA was achieved without introducing any free
parameters.  This strongly supports the applicability of DWBA for
the $^7$Li+$^7$Li system although, at first sight, the methods of
direct reaction theory seem hardly adequate for a system consisting
of such a small number of nucleons.

\bigskip
%
%\begin{footnotesize}
\textbf{Acknowledgements:}  We are grateful to Dr. E. Kwa\'sniewicz
for supplying  spectroscopic amplitudes.

\newpage

\vfill

\begin{thebibliography}{99}
%
%
\bibitem [1] {DOM87b} Domogala, G., Freiesleben, H.,
                       Nucl. Phys. {\bf A467}, 149 (1987)
%
\bibitem [2] {WIE93}   Wiebach, S., Bachmann, A.M., Brand, H., Eule, R.P.,
                         Freiesleben, H., Heyber B., Leifels, Y.,
                         Potthast, K.W., Rosenthal, P., Kamys, B.,
                         Z. Phys. {\bf A346}, 173 (1993)
%
\bibitem [3] {LEI90}   Leifels, Y., Domogala, G., Eule, R.P.,
                         Freiesleben, H., Z. Phys. {\bf A335}, 183 (1990)
%
\bibitem [4] {BAC93} Bachmann, A.M., Brand, H., Freiesleben, H.,
                       Leifels, Y., Potthast, K.W., Rosenthal, P., Kamys, B.,
                       Z. Phys. {\bf A346}, 47 (1993)
%
\bibitem [5] {GOT89} Gotzhein, I., Diploma thesis, Bochum 1989 - unpublished
%
\bibitem [6] {HUB63} Huberman, M.N., Kamegai, M., Morrison, G.C.,
                       Phys. Rev. {\bf 129}, 791 (1963)
%
\bibitem [7] {DZU64} Dzubay, T.G., Blair, J.M., Phys. Rev.
                       {\bf B134}, 586 (1964)
%
\bibitem [8] {CAR68} Carlson, R.R., Wyborny, H.W.,
                       Phys. Rev. {\bf 178}, 1529 (1968)
%
\bibitem [9] {WYB71} Wyborny, H.W., Carlson, R.R.,
                       Phys. Rev. {\bf C3}, 2185 (1971)
%
\bibitem [10] {GLU71} Glukhov, Y.A., Novatskii, B.G., Ogloblin, A.A.,
                        Sakuta, S.B., Stepanov, D.N., Chuev, V.I.,
                        Sov. J. Nucl. Phys. {\bf 13}, 154 (1971);
                        (Yad. Fiz. {\bf 13}, 277 (1971))
%
\bibitem [11] {CER74} Cerny, J., Weisenmiller, R.B., Jelley, N.A.,
                        Wilcox, K.H., Wozniak, G.J.,
                        Phys. Lett. {\bf 53B}, 247 (1974)
%
\bibitem [12] {CAR64} Carlson, R.R., McGrath, R.L.,  Norbeck, E.,
                        Phys. Rev. {\bf B136}, 1687 (1964)
%
\bibitem [13] {STR68} Stryk, R.A., Blair, J.M.,
                        Phys. Rev. {\bf 169}, 767 (1968)
%
\bibitem [14] {BOC88} Bochkarev, O.V., Korshennikov, A.A., Ku\'zmin, E.A.,
                        Mukha, I.G., Chulkov, L.V., Ya\'nkov, G.B.,
      Sov. J. Nucl. Phys. {\bf 47}, 391 (1988) (Yad. Fiz.{\bf 47}, 616 (1988))
%
\bibitem [15] {DOM87a} Domogala, G., Freiesleben, H., Hippert, B.,
              Nucl. Instr. Meth. in Phys. Research {\bf A257}, 1 (1987)
%
\bibitem [16] {TAM80} Tamura, T., Udagawa, T., Mermaz, M.C.,
                        Phys. Rep. {\bf 65}, 345 (1980);
        Program Jupiter-5, modified by Kamys, B., Rudy, Z., Wolter, H.H.,
        Zittel W., to allow for particles with non-zero spin in the
        entrance channel
%
\bibitem [17] {SAT83} Satchler, G.R., Direct Nuclear Reactions,
                         Clarendon Press, Oxford 1983
%
\bibitem [18] {OEL79} Oelert, W., Djoelis, A., Mayer-Boericke, C., Turek, P.,
                         Phys. Rev. {\bf C19}, 1747 (1979)
%
\bibitem [19] {KAM88} Kamys, B., Rudy, Z., - SQSYM computer program,
                        1988 - unpublished
%
\bibitem [20] {ENG77}  England, J.B.A., Casal, E., Garcia, A., Picazo, T.,
                       Aguilar, J., Sen Gupta, H.M.,
                       Nuclear Physics {\bf A284}, 29 (1977)
%
\bibitem [21] {HAR80} Harakeh, M.N., Van Popta, J., Saha, A., Siemssen, R.H.,
                        Nucl. Phys. {\bf A344}, 15 (1980)
%
\bibitem [22] {KWA87} Kwa\'sniewicz, E., Kisiel, J.,
                         J. of Phys. {\bf G13}, 121 (1987)
%
\bibitem [23] {KUR75} Kurath, D., Millener, D.J.,
                        Nucl. Phys. {\bf A238}, 269 (1975)
%
\bibitem [24] {KWA89} Kwa\'sniewicz, E., private information (1989)
%
\bibitem [25] {KUR73} Kurath, D., Phys. Rev. {\bf C7}, 1390 (1973)
%
\bibitem [26] {KWA86} Kwa\'sniewicz, E., Jarczyk, L.,
                         J. of Phys. {\bf G12}, 697 (1986)
%
\bibitem [27] {KWA86a} Kwa\'sniewicz, E., private information (1986)
%
\end{thebibliography}
\end{document}